\documentclass[preprint,preprintnumbers,amsmath,amssymb]{revtex4}

\usepackage{graphicx}
\usepackage{dcolumn}
\usepackage{bm}
\begin{document}

\title{ A-geometrical approach to  Topological Insulators  with defects}

\author{ D. Schmeltzer}

\affiliation{Physics Department, City College of the City University of New York \\
New York, New York 10031}


\begin{abstract}

  The  study of the propagation of electrons  with a  varying  spinor orientability is performed  using  the  coordinate  transformation method.
  
Topological Insulators  are characterized by an odd number of changes  of the orientability in the Brillouin zone.
For  defects  the change in  orientability takes place  for closed orbits  in real space.  
Both  cases  are characterized by nontrivial spin connections.

Using this method ,  we derive the form of the spin connections for topological defects in  three dimensional Topological Insulators. 
 
On the surface of a Topological Insulator, the presence an edge   dislocation gives rise to a spin connection controlled by  torsion.
 We find that  electrons propagate along  two dimensional regions  and confined  circular   contours.
We  compute  for the  edge dislocations the tunneling  density of states.
 The  edge dislocations  violates  parity symmetry resulting in a current measured  by the in-plane component of the spin on the surface.


\end{abstract}


\maketitle
\textbf{I Introduction}


The propagation of  electrons in solids is characterized by  the topological   properties  of the   the electronic band  spinors.  Topological Insulators \cite{Konig,Volkov,Gotelman,Kreutz, Mele,Kane,More,Essin,BernewigZhang,Ludwig,davidtop,ZhangField,Zhangnew} can be identified by  an odd number of changes  of the $orientability$ \cite{davidtop}  of the spinors in the  Brillouin zone. As a results non trivial spin connections  with a non- zero curvature characterized by the Chern numbers can be identified.  In  time reversal  invariant systems   one finds that for  Kramer's states  the time reversal operator  $T$  obeys   $T^2=-1$ and  one thus   the  second Chern number for    four dimensional  space is given  by $(-1)^{\nu}=-1$, where $\nu$ is an  odd number of orientability changes  \cite{Nakahara} .

Real materials are imperfect and contain topological defects such  as  dislocations \cite{Ran,Sinova},disclinations  \cite{alberto,Vozmediano} and  gauge fields induced by strain in graphene \cite{Baruch,Guinea}  ;therefore, a natural question is  to formulate the physics of Topological Insulators in the presence of such defects \cite{davidtop}.  These  topological defects  can be analyzed using  the  coordinate transformation  method  given in ref.\cite{kleinert}
which modifies the Hamiltonian for a Topological Insulator with a defect by the metric tensor and the  spin connection \cite{Pnueli,Green,Birrell,Randono,Ryu}. 

 The effect of strain fields  dislocations and disclinations plays an important role in material science and can be study using Scanning Tunneling Microscopy  ($STM$)   and  Transmission Electron Spectroscopy ($TEM$ ). Therefore we expect that the chiral metallic boundary \cite{Wu}  will be sensitive to such defects.

In this paper we will introduce the tangent space approach used in  differential geometry \cite{Nakahara,Randono,Ryu} to study propagation of electrons for a space dependent coordinate  \cite {kleinert}. We find that the continuum representation of the  edge dislocation  \cite{kleinert} generates a spin connection  \cite{Pnueli,Green,Birrell} which is controlled by the $Burger$ vector. 

Using this formulation we obtain the form of the topological insulator in three dimensions which  simplifies for the surface Hamiltonian (on the boundary). For the surface Hamiltonian we  find  that  the electronic excitations are confined  to a two-dimensional region   and to a set of   circular contours of radius $R_{g}(n)$.

 The contents of this paper is as follows: In chapter $II$  we introduce the gemetrical  method. In section $IIA$  we present the geometrical method  for  the edge dislocations and strain fields.  In section $IIB$ we consider the effects of the strain fields on the three- dimensional Topological Insulator ($TI$). The Chiral model for the boundary surface is presented in section $IIIA$. Section $IIIB$  is devoted to the derivation of the metric tensor and spin connection  for an edge dislocation \cite{kleinert}. In section $IIIC$ we identify the stable solutions. Section $IIID$ is devoted to the  stable two dimensional solutions   $n=0$ and section $IIIE$ is devoted to the  stable solution for circular  contours $n=\pm1$.
Chapter $IVA$  is devoted to the computation of the tunneling density of states. In section $IVB$  we present  results for the two dimensional  region  $n=0$. Section $IVC$ is devoted to a large number of dislocations. In section $IVD$ we compute the tunneling density of states for the circular contours $n=\pm1$. In chapter $V$ we consider the current which is given by the in-plane spin component. In section $VA$ we show that this current is zero for a $TI$. In section $VB$ we show that in the presence of an edge dislocation the parity symmetry is violated, and  current, representing the in-plane spin component, is generated. Chapter $VI$ is devoted to conclusions.

\vspace{0.2 in}

\textbf{II-The Geometrical method for dislocations and strain fields}

\vspace{0.2 in}

\textbf{A-General Considerations}

\vspace{0.1 in}

A perfect  crystal is described by    the lattice coordinates  $ \vec{r}=[x,y,z]$. For a crystal  with  a deformation , the coordinates $\vec{r}$  are replaced by 
 $ \vec{r}\rightarrow \vec{R}=\vec{r}+\vec{u}\equiv[X^{1}(\vec{r}),X^{2}(\vec{r}),X^{3}(\vec{r})]$ where $\vec{u}(\vec{r})$ is the local lattice deformation  and 
 $X^{a}$, $a=1,2,3$   is the local coordinate which  changes when we move from  one point to  another.

In a deformed crystal we introduced a set of  local vectors $e_{a}$  which are orthogonal to each other $(e_{b},e_{a})\equiv <e^{b}|e_{a}>=\delta^{b}_{a}$ and  local coordinates $X^{a}$, $a=1,2,3$. The unit vector $e_{a}$ can be represented in terms of a  \textbf{Cartesian fixed frame} space with the \textbf{ coordinate basis} $\partial_{\mu}$ ,$ \mu=x,y ,z$.  In the fixed Cartesian frame the  coordinates are given by $x^{\mu}$. Using the Cartesian basis  $\partial_{\mu}$ we  expand the deformed medium  in terms of  the local tangent vector  $e_{a}$  : $e_{a}=e^{\mu}_{a}\partial_{\mu}$  (for  the particular case where  vectors $e_{a}$ are given by  $e_{a}=\partial_{a}$,  the transformation between the two basis is  $e^{\mu}_{a}=\delta^{\mu}_{a}$).
Any vector $\vec{X} $  (in the deformed  space) can be represented in terms of the unit vectors $e_{a}$  or  the $\partial_{\mu}$ (the tangent vectors in the  Cartesian  fixed coordinates  space). The vector $\vec{X}$ can be represented in two different ways,  $\vec{X}=X^{a}e_{a}=X^{\mu}\partial_{\mu} $ (when  an index appear  twice  is understood as a summation, $X^{a}e_{a}\equiv\sum_{a=1,2,3}X^{a}e_{a}$). The dual vector  $e^{a}$ is a $one$ $form$ and can be expanded in terms of the one forms  $dx^{\mu}$. We have: $e^{a}=e^{a}_{\mu}dx$,  where $ e^{a}_{\mu}$ represents the matrix transformation $e^{a}\equiv (\partial_{\mu} X^{a})dx^{\mu}$. 
The scalar product of the components   $e^{a}_{\mu}e^{a}_{\nu}=g_{\mu,\nu}$, $e^{\nu}_{a}e^{\nu}_{b}=\delta_{a,b}$  defines the metric tensors, $g_{\mu,\nu}$  (in the Cartesian  frame ) and  $\delta_{a,b}$ in the local medium  frame.

\vspace{0.1 in}

\textbf{B-Application to the Topological insulators in three dimensions}

\vspace{0.1 in}

 The three dimensional  electronic $TI$  bands for  $Bi_{2}Se_{3}$ and $Bi_{2}Te_{3}$ can be represented using  four  projected states    \cite{Chao},  $|orbital=1,2>\otimes|spin=\uparrow,\downarrow>$   (the  Pauli matrix $\tau$ describes the orbital states and  the Pauli matrix $\sigma$ describes  the spin). The effective $h^{3D}$  Hamiltonian  in the first quantized form is given by:
\begin{equation}
h^{3D}=\hbar v_{0}[k_{y}(\sigma_{1}\otimes \tau_{1})-k_{x}(\sigma_{2}\otimes \tau_{1})+\epsilon k_{z}(\sigma_{3}\otimes \tau_{1}) + M(\vec{k})(I\otimes \tau_{3})] 
\label{h}
\end{equation}
The parameter $ M(\vec{k})$ determines if the insulator is trivial or topological. For   $Bi_{2}Se_{3}$ and $Bi_{2}Te_{3}$ the gap is inverted, namely   $M(\vec{k})=-M_{0}+B_{1}k_{z}^2+B_{2}k_{\bot}^2$ with  $M_{0}>0,B_{1}>0,B_{2}>0$ and therefore topological \cite{ZhangField,Chao,Zhangnew}.

Using the  metric tensor   $g_{\mu,\nu}$ given by the coordinate transformation ( the transformation between the two sets of coordinates - the one without the dislocation  and the second with the dislocation ) $e^{a}_{\mu}e^{a}_{\nu}=g_{\mu,\nu}$,   defines the Jacobian $\sqrt{G}$ where $G=det[g_{\mu,\nu}]$.
We find that the derivative for a spinor component  $\Psi^{(\alpha)}(\vec{r})$, $\alpha=[1=1\uparrow;2=1\downarrow;3=2\uparrow;4=2\downarrow]$   is replaced by  the $covariant$ derivative  \cite{Green}:
\begin{equation}
\nabla_{\mu}\Psi^{(\alpha)}(\vec{r})=\partial_{\mu}\Psi^{(\alpha)}(\vec{r})+\frac{1}{8}\omega^{(a,b)}_{\mu}[\hat{\Gamma}^{a},\hat{\Gamma}^{b}]^{\alpha}_{\beta}\Psi^{(\beta)}(\vec{r})
\label{vector}
\end{equation}
where $\hat{\Gamma}^{a}$ ,$a=1,2,3,4,5 $ are the matrixes:
 $\hat{\Gamma}^{1}=-\Gamma ^{2}\equiv -(\sigma_{2}\otimes\tau_{1})$; $\hat{\Gamma}^{2}=\Gamma ^{1}\equiv (\sigma_{1}\otimes\tau_{1})$; $\hat{\Gamma}^{3}=\Gamma ^{3}\equiv (\sigma_{3}\otimes\tau_{1})$; $\hat{\Gamma}^{4}=\Gamma ^{4}\equiv (I\otimes\tau_{2})$;$\hat{\Gamma}^{5}=\Gamma ^{5}\equiv (I\otimes\tau_{3})$.
 
The $spin$ $connection$   $\omega^{a,b}_{\mu}$  determines the covariant derivative  \cite{Green}  is given in terms of the tangent vectors $ e^{a}_{\mu}$: $e^{a}_{\mu}=\partial_{\mu}X^{a}(\vec{r})$; \hspace{0.1 in} $a=1,2,3$ ; \hspace{0.1 in}  $\mu=x,y,z$.
\begin{eqnarray} &&\omega^{a,b}_{\mu}=\frac{1}{2}e^{\nu,a}(\partial_{\mu}e^{b}_{\nu}-\partial_{\nu}e^{b}_{\mu})-
\frac{1}{2}e^{\nu,b}(\partial_{\mu}e^{a}_{\nu}-\partial_{\nu}e^{a}_{\mu})\nonumber\\&&
-\frac{1}{2}e^{\rho,a}e^{\sigma,b}(\partial_{\rho}e_{\sigma, c}-\partial_{\sigma}e_{\rho, c})e^{c}_{\mu}
\end{eqnarray}
We notice the asymmetry   between $e^{\nu,a}$ and  $e_{a,\nu}$: 
$e^{\nu,a}\equiv g^{\nu,\lambda}e^{a}_{\lambda}$ and  $e_{a,\nu}\equiv \delta_{a,b} e^{b}_{\nu}$.
As a result the  Hamiltonian in eq.$(1)$ in the second quantized form  is replaced by:
\begin{eqnarray}
&&H^{(3D)} = \hbar v_{0}\int\,d^{3}r \sqrt{G}[\Psi^{\dagger}(\vec{r})[e^{\mu}_{a}\hat{\Gamma}^{a} (-i\nabla_{\mu}) -E_{F}(I\otimes I)+ \hat{\Gamma}^{5}(-M_{0})]\Psi(\vec{r})\nonumber\\&&+B_{1}g^{\mu,\nu} (\nabla_{\mu}\Psi^{\dagger}_{1}(\vec{r}))( \nabla_{\nu}\Psi_{1} (\vec{r}))- B_{1}g^{\mu,\nu} (\nabla_{\mu}\Psi^{\dagger}_{2}(\vec{r}) \nabla_{\nu}\Psi_{2})]\nonumber\\&&
\end{eqnarray}

where $e^{\mu}_{a}\hat{\Gamma}^{a}=\sum_{a}e^{\mu}_{a}\hat{\Gamma}^{a}\equiv \hat{\Gamma}^{\mu}(\vec{r})$, 
 $[\hat{\Gamma}^{\mu}(\vec{r})\hat{\Gamma}^{\nu}(\vec{r})+\hat{\Gamma}^{\mu}(\vec{r})\hat{\Gamma}^{\nu}(\vec{r})]=2g^{\mu,\nu}(\vec{r})$ , $det[g^{\mu,\nu}(\vec{r})]\equiv G$ and  $\nabla_{\mu}$ is the covariant derivative  given in terms of the spin connection given in equation $(2)$:
 $\nabla_{\mu}\Psi^{(\alpha)}(\vec{r})=\partial_{\mu}\Psi^{(\alpha)}(\vec{r})+\frac{1}{8}\omega^{(a,b)}_{\mu}[\hat{\Gamma}^{a},\hat{\Gamma}^{b}]^{\alpha}_{\beta}\Psi^{(\beta)}(\vec{r})$ 

\vspace{0.1 in}

\textbf{C-The Mechanical strain effect on  $H^{(3D)}$}

\vspace{0.1 in}

From the work of  \cite{young} we learn that the effect of the strained field is different on $Bi_{2}Se_{3}$ than   on $Bi_{2}Te_{3}$.  In $Bi_{2}Se_{3}$ the $compressive$ strain decreases the  Coulombic gap while increasing  the inverted gap strength induced by  the spin-orbit  interaction.
We will use the result in equation $(4)$ to analyze the effect of strain.
 The strain field $\epsilon_{i,j}$ (symmetric in $i,j$) is related to the stress field $\sigma_{i,j}$ and elastic  stiffness  $Lame$ constant $\lambda$ and $\mu$: $\sigma_{i,j}=\lambda\delta_{i,j}\epsilon_{k,k}+2\mu\epsilon_{i,j}$.
Applying a constant stress $\sigma_{i,j}$ one can determine the value of the constant strain field $\epsilon_{i,j}$ which is  related to the  tangent vectors  $e^{i}_{j}\equiv \delta_{i,j}+\epsilon_{i,j}$. In the present case the spin connection and the  \textbf{Christofel} tensor   vanish.  The metric tensor  $g_{i,j}$ is given by :$g_{i,j}= \delta_{i,j}+2 \epsilon_{i,j}$. 
Using this formulation we can investigate the effect of the stress on the $Bi_{2}Se_{3}$ at the $\Gamma$ point  $\vec{k}=0$.
The TI Hamiltonian given in eq.$(4)$ ,$M(\vec{k})=-M_{0}+B_{1}k^{2}_{z}+B_{2}(k^{2}_{x}+ k^{2}_{y})$  with the inverted case $M_{0}>0$  \cite{Zhangnew} .
The Hamiltonian  in eq. $(4)$ is replaced by: 
\begin{eqnarray}
&&H^{(3D-strain)}=\hbar v_{0}\int\,d^{3}r \sqrt{G}[\Psi^{\dagger}(\vec{r})[\hat{\Gamma}^{a}(\delta_{\mu,a}+\epsilon_{\mu,a})(-i\partial_{\mu}))+ \hat{\Gamma}^{4}(-M_{0})+ B (1-2\epsilon_{\mu,\nu})\partial_{\mu}\hat{\Gamma}^{4}\partial_{\nu}]\Psi(\vec{r})\nonumber\\&&\approx \hbar v_{0}\int\,d^{3}r \sqrt{G}[\Psi^{\dagger}(\vec{r})[\hat{\Gamma}^{a}[(\delta_{\mu,a}(1+<\epsilon>)(-i\partial_{\mu}))]+ \hat{\Gamma}^{4}(-M_{0})+ B (1-2<\epsilon>\delta_{\mu,\nu} )\partial_{\mu}\hat{\Gamma}^{4}\partial_{\nu}]\Psi(\vec{r})\nonumber\\&&
\end{eqnarray}
In equation  $(5)$   we have  used the average strain field  $<\epsilon>$, $<\epsilon>\equiv\frac{\epsilon_{1,1}+\epsilon_{2,2}+\epsilon_{3,3}}{3}$.
We replace the spinor field    $\Psi(\vec{r})$ by $\Psi(\vec{r})\sqrt{(1+<\epsilon>)}\equiv\hat{\Psi}(\vec{r})$. As a result we obtain:
\begin{eqnarray} 
&&H^{(3D-strain)}\approx\hbar v_{0}\int\,d^{3}r \sqrt{G}[\hat{\Psi}^{\dagger}(\vec{r})\hat{\Gamma}^{\mu}(-i\partial_{\mu})+ \hat{\Gamma}^{4}\frac{(-M_{0})}{(1+<\epsilon>)}+ B \frac{(1-2<\epsilon>)}{1+<\epsilon>} \partial_{\mu}\hat{\Gamma}^{4}\partial_{\mu}]\hat{\Psi}(\vec{r})\nonumber\\&&
\end{eqnarray}
 
For the compressive case    $<\epsilon>$ is negative,   $<\epsilon> \equiv-|<\epsilon>|$ .
As a result we observe that the inverted gap is enhanced $\frac{|M_{0}|}{(1+<\epsilon>)}=\frac{|M_{0}|}{(1-|<\epsilon>|)}> |M_{0}|$. 

In the same way we can show that the Coulomb interaction is reduced:
We introduce the Hubbard Stratonovici field  $a_{0}$  to describe the Coulomb  interactions.
\begin{eqnarray}
&&H^{e-e}=\int\,d^{3}r \sqrt{G}[ I (-e)\cdot a_{0}\Psi^{\dagger}(\vec{r})\Psi(\vec{r}) +\frac{(1-2\epsilon_{\mu,\nu})}{2}a_{0}\partial_{\mu}\partial_{\nu} a_{0}]\nonumber\\&& \approx 
\int\,d^{3}r \sqrt{G}[ I (-e)\cdot a_{0}\Psi^{\dagger}(\vec{r})\Psi(\vec{r}) +\frac{(1-2<\epsilon>)}{2}a_{0}\partial_{\mu}\partial_{\nu} a_{0} ]\nonumber\\&&
\end{eqnarray}
Next we rescale  $a_{0}= \frac{ A_{0}}{\sqrt{(1-2<\epsilon>)}}$ and obtain:
\begin{equation}
H^{e-e}\approx \int\,d^{3}r \sqrt{G}[ I \frac{(-e)}{\sqrt{1-2<\epsilon>}} A_{0}\Psi^{\dagger}(\vec{r})\Psi(\vec{r}) +A_{0}\partial_{\mu}\partial_{\mu} A_{0} ]
\label{coulomb}
\end{equation}
We observe that  for  the compressive case  the effective charge $e_{eff.}\equiv\frac{(-e)}{\sqrt{1-2<\epsilon>}}=\frac{(-e)}{\sqrt{1+2|<\epsilon>|}}$  is reduced   and therefore the Coulomb gap decreases, while at the same time the  inverted gap  increases,  $\frac{|M_{0}|}{(1+<\epsilon>)}=\frac{|M_{0}|}{(1-|<\epsilon>|)}> |M_{0}|$ in qualitative agreement with  \cite{young}.

\vspace{0.2 in}
 
\textbf{III-The chiral metal with an edge dislocation}

\vspace{0.2 in}

\textbf{A-Description of   the Chiral model}

\vspace{0.1in}

The low energy  Hamiltonian for the bulk $3D$ $TI$ in the $Bi_{2}Se_{3}$ family was  shown to behave on the boundary  surface  (the $x,y$- plane) as  a two dimensional   chiral metal  \cite{nature} .
\begin{eqnarray}
&&H= \int\,d^{2}r \Psi^{\dagger}(\vec{r})[h^{T.I}-\mu]\Psi(\vec{r})]\equiv \hbar v_{F}\int\,d^{2}r \Psi^{\dagger}(\vec{r})[i\sigma^{1}\partial_{y} -i\sigma^{2}\partial_{x}-\mu ]\Psi(\vec{r})\nonumber\\&&
\end{eqnarray}
$h^{T.I}=\hbar v_{F}[i\sigma^{1}\partial_{y} -i\sigma^{2}\partial_{x}]$
is the   chiral Dirac Hamiltonian in the first quantized language. $v_{F}\approx  5\cdot10^{5} \frac{m}{sec}$ is the Fermi velocity, $\sigma$ is the Pauli matrix describing the electron spin and $\mu$ is the chemical potential  measured relative to the Dirac $\Gamma$ point.
The  Hamiltonian  for the  two dimensional surface $L\times L$ describes well the excitations smaller than the bulk gap of the $3D$ $TI$ at $0.3$ $eV$. Moving away from  the $\Gamma$ point, the Fermi velocity becomes momentum dependent; therefore,  we will introduce a  momentum  cut off  $\Lambda $  to restrict the validity of the Dirac model.
The chiral Dirac model in the Bloch representation   takes the form: $h=\hbar v_{F}(  \vec{K}\times\vec{\sigma})\cdot\hat{z}\equiv  \hbar v_{F}(-\sigma^{1}k_{y}+\sigma^{2}k_{x})$ 
 The  eigen-spinors for this Hamiltonian are :
$|u(\vec{K})>=[|u_{\uparrow}(\vec{K})>,|u_{\downarrow}(\vec{K})>]^{T}=|\vec{K}>\otimes[1,i e^{i\chi(k_{x},k_{y})}]^{T}$  where  $\chi(k_{x},k_{y})=tan^{-1}(\frac{k_{y}}{k_{x}})$  is the spinor phase and  $\epsilon=\hbar v_{F}\sqrt{k^{2}_{x}+k^{2}_{y}}$ is the eigenvalue  for particles .  For holes we have the eigenvalue   $\epsilon=-\hbar v_{F}\sqrt{k^{2}_{x}+k^{2}_{y}}$  and eigenvectors  $|v(\vec{K})>=[|v_{\uparrow}(\vec{K})>,|v_{\downarrow}(\vec{K})>]^{T}= |\vec{K}>\otimes[-1, i e^{i\chi(k_{x},k_{y})}]^{T}$.
The \textbf{chirality}  operator is defined  in terms of the chiral phase  $\chi(k_{x},k_{y})$: 
\begin{equation}
(\vec{\sigma}\times \frac{\vec{K}}{ |\vec{K}|})\cdot\hat{z}\equiv \sin[\chi(k_{x},k_{y})]\sigma^{1}-\cos[\chi(k_{x},k_{y})]\sigma^{2}
\label{eqchirality}
\end{equation}
The chirality operator  takes the eigenvalue  $-$ (counter-clockwise) for  particles 
$[\sin(\chi(k_{x},k_{y}))\sigma^{1}-\cos(\chi(k_{x},k_{y}))\sigma^{2}]|\vec{K}>\otimes[1,i e^{i\chi(k_{x},k_{y})}]^{T}=-|\vec{K}>\otimes[1,i e^{i\chi(k_{x},k_{y})}]^{T}$
and  $+$ (clockwise) for holes
$[\sin(\chi(k_{x},k_{y}))\sigma^{1}-\cos(\chi(k_{x},k_{y}))\sigma^{2}]|\vec{K}>\otimes[-1,i e^{i\chi(k_{x},k_{y})}]^{T}=|\vec{K}>\otimes[-1,i e^{i\chi(k_{x},k_{y})}]^{T}$.


\vspace{0.1 in}

\textbf{B-The  effect of edge dislocation on a two dimensional  chiral surface Hamiltonian}

\vspace{0.1 in}

 We use the notation $ x^{\mu}$ ,$\mu=x,y$   and $X^{a}$ ,$a=1,2$ to describe  the media with dislocations.
For an edge dislocation in the $x$ direction   the $Burger$ vector $B^{(2)}$  is in the $y$ direction . The value of the burger vector   $B^{(2)}$  is given by   the   shortest translation  lattice  vector   in the $y$ direction.  (For  the  $TI$  $Bi_{2}Se_{3}$ the length of the vector  $B^{(2)}$ is $5$ times  the inter atomic  distance ).
Following \cite{kleinert} we introduce  the \textbf{coordinate transformation for an edge dislocation}: $\vec{r}=(x,y)\rightarrow  [X(\vec{r})=x,Y(\vec{r})=y+\frac{B^{(2)}}{2\pi}\tan^{-1}(\frac{y}{x})]$ with the core of the  dislocation centered at $\vec{r}=(0,0)$.  
  The  matrix elements  fields $ e^{a}_{\mu}$ for the edge dislocation is given by  :
\begin{equation}
e^{a}_{\mu}=\partial_{\mu}X^{a}(\vec{r}); \hspace{0.1 in} a=1,2 ; \hspace{0.1 in}  \mu=x,y
\label{ea}
\end{equation}
 We  express   the Burger vector  in terms of the the partial derivatives  with respect the coordinates $a=1,2$ in the dislocation frame and $\mu=x,y$ for the fixed Cartesian frame \cite{kleinert}:
\begin{equation} 
 \partial_{x}e^{2}_{y}-\partial_{y}e^{2}_{x} = B^{(2)}\delta^{2}(\vec{r})
\label{stokes}
\end{equation} 
Using Stokes theorem, we  replace the line integral $\displaystyle\oint dx^{\mu}e^{2}_{\mu}(\vec{r})$  by the surface integral   $\int\int dx^{\mu}
 dx^{\nu}[\partial_{x}e^{2}_{y}-\partial_{y}e^{2}_{x}]  $. For a system with zero $curvature$   and non zero $torsion$  $T^{(2)}_{\mu,\nu}$ we find that  the  surface torsion tensor integral $\int\int dx^{\mu}
 dx^{\nu}T^{(2)}_{\mu,\nu}$  is equal to  $\int\int dx^{\mu}
 dx^{\nu}[\partial_{x}e^{2}_{y}-\partial_{y}e^{2}_{x}]  $,  and therefore both   integrals are equal  to  the Burger vector.
\begin{eqnarray}
&&\displaystyle\oint dx^{\mu}e^{2}_{\mu}(\vec{r})= \int\int dx^{\mu} dx^{\nu}[\partial_{\mu}e^{2}_{\nu}-\partial_{\nu}e^{2}_{\mu}]= B^{(2)};\nonumber\\&&
\int\int dx^{\mu}
 dx^{\nu}T^{(2)}_{\mu,\nu}=\int\int dx^{\mu} dx^{\nu}[\partial_{\mu}e^{2}_{\nu}-\partial_{\nu}e^{2}_{\mu}]= B^{(2)};\nonumber\\&&
\end{eqnarray}
where $dx^{\mu}dx^{\nu}$ represents the surface element.
The  tangent components  $e^{a}_{\mu}$  can be expressed in terms  of the Burger  vector  density $ B^{(2)}\delta^{2}(\vec{r})$  \cite{kleinert} :
\begin{eqnarray}
&&e^{2}_{x}=(\frac{B^{(2)}}{2\pi})\frac{y}{(x^2+y^2)};\hspace{0.3 in} e^{2}_{y}=1-(\frac{B^{(2)}}{2\pi})\frac{x}{(x^2+y^2)} \nonumber\\&&e^{1}_{x}=1 ;\hspace{0.4 in}e^{1}_{y}=0
\end{eqnarray} 
Using the tangent components, we obtain the metric tensor $g_{\mu,\nu}$. 
\begin{equation}
e^{a}_{\mu}e^{a}_{\nu}\equiv e^{1}_{\mu}e^{1}_{\nu}+ e^{2}_{\mu}e^{2}_{\nu}=g_{\mu,\nu}(\vec{r});\hspace{0.1 in}
e^{a}_{\mu}e^{b}_{\mu}\equiv e^{a}_{x}e^{b}_{x}+ e^{a}_{y}e^{b}_{y}=\delta_{a,b}
\label{ec}
\end{equation} 
The inverse of the metric tensor  $g_{\mu,\nu}(\vec{r})$ is the tensor  $g^{\nu,\mu}(\vec{r})$
defined trough the equation $g_{\mu,\tau}(\vec{r})g^{\tau,\nu}(\vec{r})=\delta_{\mu}^{\nu}$.
Using the tangent vectors,  we find   $to$ $first$ $order$ in the Burger vector the metric tensor $g_{\mu,\nu}$ and the  Jacobian transformation $\sqrt{G}$:
\begin{equation}
g_{x,x}=1;\hspace {0.1 in} g_{x,y}=\frac{B^{(2)}}{2\pi}\frac{y}{x^{2}+y^{2}};\hspace{0.1 in}  g_{y,y}=1-\frac{B^{(2)}}{2\pi}\frac{y}{x^{2}+y^{2}}; \hspace{0.1 in} g_{y,x}=0; \hspace{0.1 in}
G=det[g_{\mu,\nu}]=1-\frac{B^{(2)}}{2\pi}\frac{y}{x^{2}+y^{2}}
\label{metric}
\end{equation}
The inverse tensor is given by:$g^{x,x}\approx 1$, $g^{x,y}=g^{y,x}=-\frac{B^{(2)}}{2\pi}\frac{y}{x^{2}+y^{2}}$, $g^{y,y}=1+\frac{B^{(2)}}{\pi}\frac{x}{x^{2}+y^{2}}$.
Using the inverse tensor $ g^{\mu,\nu}$ we obtain the inverse matrix $e^{\mu}_{a} $ which is given by:
\begin{equation}
e^{\mu}_{a}=e_{a,\nu}g^{\nu,\mu}=(\delta_{a,b}e^{b}_{\nu})g^{\nu,\mu}=e^{a}_{\nu}g^{\nu,\mu}
\label{trans}
\end{equation}

Using the  components $e^{\mu}_{a} $  we compute the  the transformed Pauli matrices.
The Hamiltonian in the absence of the edge  dislocation is given by $h^{T.I.}=i \gamma^{a}\partial_{a}\equiv \sum_{a=1,2} i\gamma^{a}\partial_{a}$ where the Pauli matrices  are given by  $\gamma^{1}=-\sigma^{2}$ , $\gamma_{2}=\sigma_{1}$  and $\gamma^{3}=\sigma^{3}$. (We will use the convention that when an index appears twice we  perform a summation  over this index.)  In the presence of the   edge dislocation, the term $\gamma^{a}\partial_{a}$  must   be expressed in terms of the Cartesian   fixed coordinates  $\mu=x,y$. As a result, the spinor $\Psi(\vec{r})$    transforms accordingly to the $SU(2)$ transformation . If $\widetilde{\Psi}(\vec{R})$ is the spinor for the deformed lattice, it can be related with the help of an $SU(2)$ transformation to the spinor  $\Psi(\vec{r})$ in the  undeformed lattice:  $\widetilde{\Psi}(X,Y)=e^{-i\frac{\delta\varphi(x,y)}{2}\sigma^{3}}\Psi(x,y)$ .  Where   $\delta\varphi(x,y)$ is the rotation angle between the two  set of coordinates:
$\delta\varphi(x,y)= tan^{-1}(\frac{Y}{X})-tan^{-1}(\frac{y}{x})$. Using the relation between the coordinates  $X=x$, and $Y=y+\frac{B^{(2)}}{2\pi}tan^{-1}(\frac{y}{x})$ with the singularity at $x=y=0$  gives us that the derivative of the  phase which is a delta function, $\partial_{x}\delta\varphi(x,y)=-\partial_{y}\delta\varphi(x,y)\propto \delta^{2}(x,y)$. Combining the  transformation of the derivative with the $SO(2)$ rotation in the plane,  we obtain  the form of the chiral Dirac equation in the Cartesian space  (see Appendix A) given in terms of the  $spin$ $connection$  $\omega_{\mu}^{1, 2}$   \cite{Nakahara}:
\begin{equation}
i\gamma_{a}\partial_{a}\widetilde{\Psi}(\vec{R})= i\delta_{a,b}\gamma^{b}\partial_{a}\widetilde{\Psi}(\vec{R})=i\gamma^{a}e^{\mu}_{a}[\partial_{\mu}+\frac{1}{4}[\gamma^{b},\gamma^{c}]\omega_{\mu}^{b c}]\Psi(\vec{r})
\label{trans}
\end{equation}
The Hamiltonian $h^{T.I.}\rightarrow h^{edge}$  is transformed to the dislocation edge Hamiltonian  with the explicit form given by:
\begin{eqnarray}
&&h^{edge}=i\sigma^{1}\partial_{2}-i\sigma^{2}\partial_{1}=
i\sigma^{1}e^{\mu}_{2}[\partial_{\mu}+\frac{1}{8}[\sigma^{1},\sigma^{2}]\omega ^{1,2}_{\mu}]-i\sigma^{2}e^{\mu}_{1}[\partial_{\mu}+\frac{1}{8}[\sigma^{1},\sigma^{2}]\omega ^{1,2}_{\mu}]\nonumber\\&&=
i(\sigma^{1}e^{\mu}_{2}-\sigma^{2}e^{\mu}_{1})
(\partial_{\mu}+\frac{1}{8}[\sigma^{1},\sigma^{2}]\omega ^{1,2}_{\mu})
\nonumber\\&&
\end{eqnarray}
To first  order in the Burger vector we find : $\omega_{x}^{1 2}=- \omega_{x}^{2 1 }=0$ and $- \omega_{y}^{2 1 }=\omega_{y}^{1, 2}= -\frac{B^{(2)}}{2}\delta^{2}(\vec{r}) $, see eqs. $(72-74)$  in Appendix A.
\begin{equation}
h^{edge}\approx i\sigma^{1}(\partial_{y} -\frac{i}{2} \sigma^{3}B^{(2)}\delta^{2}(\vec{r})) -i\sigma^{2}\partial_{x}
\label{Burger}
\end{equation}
In the second quantized form  the chiral Dirac Hamiltonian in the presence of an edge dislocations  is given by :
\begin{eqnarray}
&&H^{edge}\approx \int\,d^{2}r \sqrt{G}\Psi^{\dagger}(\vec{r})[h^{edge}-\mu]\Psi(\vec{r})\nonumber\\&&\equiv \hbar v_{F}\int\,d^{2}r \sqrt{G}\Psi^{\dagger}(\vec{r})[i\sigma^{1}(\partial_{y} -\frac{i}{2} \sigma^{3}B^{(2)}\delta^{2}(\vec{r}))-i\sigma^{2}\partial_{x} -\mu ]\Psi(\vec{r})\nonumber\\&&
\end{eqnarray}
$h^{edge}$
is the Hamiltonian in the first quantized language, $\mu$ is the chemical potential and $\Psi(\vec{r})=[\Psi_{\uparrow}(\vec{r}),\Psi_{\downarrow}(\vec{r})]^{T}$ is the two component spinor field.

\vspace{0.1 in}

\textbf{C- The Identification of the physical contours for the  edge  Hamiltonian $h^{edge}$}

\vspace{0.1 in}

In order to identify the solutions, we will use the complex representation.
The coordinates in   the   complex representation are given by,
$z=\frac{1}{2}(x+iy)$,\hspace{0.1 in} $\overline{z}=\frac{1}{2}(x-iy)$, \hspace{0.1 in}
$\partial_{z}=\partial_{x}-i\partial_{y}$, \hspace{0.1 in} $\partial_{\overline{z}}=\partial_{x}+i\partial_{y}$. In this representation    the two dimensional  delta function  $\delta^{2}(\vec{r})$ is given by  $\delta^{2}(\vec{r})\equiv\frac{1}{\pi} 
\partial_{z}(\frac{1}{ \overline{z}})=\frac{1}{\pi} 
\partial_{\overline{z}}(\frac{1}{z})$ \cite{Conformal,Nair}.
We will  use the edge Hamiltonian  $h^{edge}   $
and will compute the eigenfunctions  $u_{\epsilon}(z,\overline{z})=[U_{\epsilon\uparrow}(z,\overline{z}),U_{\epsilon\downarrow}(z,\overline{z})]^{T}$ and $v_{\epsilon}(z,\overline{z})=V_{\epsilon\uparrow}(z,\overline{z}),V_{\epsilon\downarrow}(z,\overline{z})]^{T}$.
The eigenvalue equation is given by:
\begin{eqnarray}
&&\epsilon U_{\epsilon\uparrow}(z,\overline{z})=-[\partial_{z}+(\frac{B^{(2)}}{ \sqrt{2}\pi})\partial_{z}(\frac{1}{ \overline{z}})] U_{\epsilon\downarrow}(z,\overline{z})\nonumber\\&&
 \epsilon U_{\epsilon \downarrow}(z,\overline{z})=[\partial_{\overline{z}}+ (\frac{ B^{(2)}}{\sqrt{2}\pi})\partial_{\overline{z}}(\frac{1}{z})] U_{\epsilon\uparrow}(z,\overline{z})\nonumber\\&&
\end{eqnarray} 
The eigenfunctions  $u_{\epsilon}(z,\overline{z})$ and $v_{\epsilon}(z,\overline{z})$ can be written with the help of   a singular  matrix $ M(z,\overline{z})$ \cite{Ezawa} :
\begin{equation*}
u_{\epsilon}(z,\overline{z})=M(z,\overline{z})\hat{F}_{\epsilon}(z,\overline{z})\equiv\left[\begin{array}{rrr}
e^{-\frac{B^{(2)}}{2\pi}(\frac{1}{ z})}& 0 \\
0 & e^{-\frac{B^{(2)}}{2\pi}(\frac{1}{ \overline{z}})}\\
\end{array}\right]
\left(\begin{array}{cc}F_{\epsilon\uparrow}(z,\overline{z})\\F_{\epsilon\downarrow}(z,\overline{z})\end{array}\right)
\end{equation*}
 ($F_{\epsilon}(z,\overline{z})$ and $F_{-\epsilon}(z,\overline{z})$  are the   transformed eigenfunctions for $\epsilon>0$ and   $\epsilon <0$  respectively .) In terms of the transformed spinors 
the eigenvalue equation $h^{edge}(z,\overline{z})u_{\epsilon}(z,\overline{z})=\epsilon u_{\epsilon}(z,\overline{z})$ and $F_{\epsilon\downarrow}(z,\overline{z})$ becomes: 
\begin{equation*}
\epsilon\left(\begin{array}{cc}F_{\epsilon\uparrow}(z,\overline{z})\\F_{\epsilon\downarrow}(z,\overline{z})\end{array}\right)=\left[\begin{array}{rrr}
I(z,\overline{z})& 0 \\
0 & (I(z,\overline{z})^{*}\\
\end{array}\right]\left[\begin{array}{rrr}
-\partial_{z} & 0 \\
0 & \partial_{\overline{z}} \\
\end{array}\right]\left(\begin {array}{cc}F_{\epsilon\uparrow}(z\overline{z})\\F_{\epsilon\downarrow},(z,\overline{z})\end{array}\right)
\end{equation*}
where  $I(z,\overline{z})= e^{-\frac{B^{(2)}}{2\pi}(\frac{\overline{z}-z}{ z \overline{z}})}\equiv e^{2\frac{B^{(2)}}{\pi}(\frac{ iy}{x^2+y^2})}$ , $(I(z,\overline{z}))^{*}= e^{2\frac{B^{(2)}}{\pi}(\frac{-iy}{x^2+y^2})}$, $|I(z,\overline{z})|=1$. 
We search for zero modes  $\epsilon=0$ and find  :
\begin{equation}
\partial_{z}F_{\epsilon\downarrow}(z,\overline{z})=0\hspace{0.5 in}
\partial_{\overline{z}}F_{\epsilon\uparrow}(z,\overline{z})=0
\label{zero}
\end{equation}
The solutions are given by the holomorphic   representation 
$F_{\epsilon=0\uparrow}(z,\overline{z})=f_{\uparrow}(z)$  and    the anti-holomorphic function   $F_{\epsilon=0\downarrow}(z,\overline{z})=f_{\downarrow}(\overline{z})$.
The  zero mode eigenfunctions are  given by :
\begin{equation}
u_{\epsilon=0,\uparrow}(z)=e^{-\frac{ B^{(2)}}{2\pi}(\frac{1}{z})} f_{\uparrow}(z),\hspace{0.5 in}
u_{\epsilon=0,\downarrow}(\overline{z})=e^{-\frac{ B^{(2)}}{2\pi}(\frac{1}{\overline{z}})}f_{\downarrow}(\overline{z})
\label{eigen}
\end{equation}
Due to the presence of the essential singularity at $z=0$ it is not possible to find  analytic functions  $f_{\uparrow}(z)$  and   $f_{\downarrow}(\overline{z})$   which  vanish fast enough around   $z=0$ such that $\int d^{2}z (u_{\epsilon=0,\lambda}(z))^{*} u_{\epsilon=0,\lambda}(z)<\infty $. Therefore, we conclude that  \textbf{zero mode solution} does not exists.    
The only way to remedy the problem is to allow for states with finite energy. 

 In the next step  we  look for finite energy states.
We perform a coordinate transformation :
\begin{equation}
z\rightarrow W[z,\overline{z}];\hspace{0.5 in}  \overline {z}\rightarrow\overline{W}[z,\overline{z}]
\label{transform}
\end{equation}
We  demand that the transformation is conformal and preserve the orientation. This restricts the transformations to   holomorphic and anti-holomorphic functions \cite{Conformal}. This means that  we  have the conditions    $\partial_{\overline{z}} W[z,\overline{z}]=0 $ and $\partial_{z} \overline{W}[z,\overline{z}]=0 $. As a result  we obtain   $W[z,\overline{z}]=W[z]$ and  $\overline{W}[z,\overline{z}]=\overline{W}[\overline{z}]$, which obey  the eigenvalue equations: 
\begin{eqnarray}
&&\epsilon F_{\epsilon\uparrow}(W,\overline{W})=-\partial_{W}F_{\epsilon\downarrow}(W,\overline{W})\nonumber\\&&
\epsilon F_{\epsilon\downarrow}(W,\overline{W})= \partial_{\overline {W}}F_{\epsilon\uparrow}(W,\overline{W})\nonumber\\&&
\end{eqnarray}
This implies the conditions $\frac{dW[z]}{dz}=(I(z,\overline{z}))^{*}$ and  $\frac{d\overline{W}[\overline{z}]}{dz}=I(z,\overline{z})$. Since  $I(z,\overline{z})$  is neither holomorphic or anti-holomorphic  and satisfy    $|I(z,\overline{z})|=1$, the only   solutions for $W[z]$ and $\overline{W}[\overline{z}]$ must obey      $I(z,\overline{z})=1$: 
\begin{equation}
I(z,\overline{z})\equiv e^{2\frac{B^{(2)}}{\pi}(\frac{ iy}{x^2+y^2})}=
e^{i 2\pi n};\hspace{0.2 in} n=0,\pm 1,\pm 2....
\label{solution}
\end{equation} 
For   $I(z,\overline{z})\neq 1$  one obtains solutions which are unstable .    The stable solutions  will be given by a  one parameter $s$   curve  ($s$ is the length of the curve)  $\vec{r}(s)\equiv[x(s),y(s)]$      which obey the equation  $I(z,\overline{z})=1$. 
 The curve $\vec{r}(s)$  allows  us to  define the $tangent$ $\vec{t}(s)$  and the  $normal$ vectors   $\vec{N}(s)$. This allows us  to introduce   a two- dimensional region in the vicinity of the contour  of $\vec{r}(s)\rightarrow  \vec{R}(s,u)=\vec{r}(s)+u \vec{N}(s)$.

\vspace{0.1 in}

\textbf{IID- The wave function for the edge dislocation-the $n=0$ contour}

\vspace{0.1 in}


\vspace{0.1 in}

 The condition $I(z,\overline{z})=e^{2\frac{B^{(2)}}{\pi}(\frac{ iy}{x^2+y^2})} =1$   for   $n=0$ is satisfied  for  $y=0$ and large  value of $y$  which obey    $2\frac{B^{(2)}}{\pi}(\frac{ y}{x^2+y^2})<<1$ .  The  values of $y$   which  satisfy   this conditions are restricted to   $I(z,\overline{z})=e^{2\frac{B^{(2)}}{\pi}(\frac{ iy}{x^2+y^2})}\approx 1$.
This  condition is satisfied  for values of  $y$  in the range:
\begin{equation} 
2\frac{B^{(2)}}{\pi}(\frac{ y}{x^2+y^2})\leq \eta <\frac{\pi}{4}<1
\label{eta}
\end{equation}
We introduce the radius   $R_{g}=\frac{B^{2}}{2\pi^2}$  and find that the condition   $I(z,\overline{z})\approx1$ gives rise to the  equation for $y$. The solution   is given by  $ x^2+ (y\pm \frac{2\pi}{\eta}R_{g})^2 =( \frac{2\pi}{\eta}R_{g})^2$. 
Therefore,  for $|y|>|d|\geq(\frac{2\pi}{\eta}) 2R_{g}> 2R_{g}$ we have   $I\approx1$ which corresponds to   a free particle  eigenvalue equations.
\begin{eqnarray}
&&\epsilon F_{\epsilon\uparrow}(x,y)=e^{\frac{ B^{(2)}}{\pi}  \frac{i2y}{(x^2+y^2)}}[-\partial_{x} +i\partial_{y}]F_{\epsilon\downarrow}(x,y)\nonumber\\&&
\approx  [-\partial_{x} +i\partial_{y}]F_{\epsilon\downarrow}(x,y);\nonumber\\&& \epsilon F_{\epsilon\downarrow}(x,y)=  e^{\frac{ B^{(2)}}{\pi}  \frac{-i2y}{(x^2+y^2)}}[\partial_{x}+i\partial_{y}]F_{\epsilon\uparrow}(x,y)\nonumber\\&&\approx[\partial_{x}+i\partial_{y}]F_{\epsilon\uparrow}(x,y)\nonumber\\&&
\end{eqnarray}
For $|y|>d$  the eigenfunctions are given by:
$U_{\epsilon,\uparrow}(x,y)=e^{-\frac{B^{(2)}}{2\pi}(\frac{1}{x+iy})}F_{\epsilon,\uparrow}(x,y)$, $U_{\epsilon,\downarrow}(x,y)=e^{-\frac{B^{(2)}}{2\pi}(\frac{1}{x-iy})}F_{\epsilon,\downarrow}(x,y)$ where $F_{\epsilon\uparrow}(x,y)$ and $F_{\epsilon\downarrow}(x,y)$ are  the eigenfunctions of  equation (21).   The  envelope functions  $e^{-\frac{ B^{(2)}}{2\pi}(\frac{1}{x+iy})}$, $e^{-\frac{ B^{(2)}}{2\pi}(\frac{1}{x-iy})}$  which  multiply  the wave functions   $F_{\epsilon\uparrow}(x,y)$ , $F_{\epsilon\downarrow}(x,y)$  impose  vanishing  boundary  conditions for the  eigenfunctions $U_{\epsilon,\downarrow}(x,y)$ and $U_{\epsilon,\uparrow}(x,y)$ at  $y\rightarrow\pm \infty$ .   therefore, we   demand that the  eigenfunctions $U_{\epsilon,\uparrow}(x,y)$, $U_{\epsilon,\downarrow}(x,y)$ should   vanish  at the boundaries  $y=\pm\frac{L}{2}$.
Since the multiplicative envelope functions for opposite spins is  complex conjugate to each other we have to make the choice that one of the spin components vanishes at one side and the other component at the opposite side. Two possible choices   can be made: 

$U_{\epsilon,\uparrow}(x,y=\frac{L}{2})\equiv e^{-\frac{ B^{(2)}}{2\pi}(\frac{1}{x+i\frac{L}{2}})}F_{\epsilon\uparrow}(x,\frac{L}{2})=  U_{\epsilon,\downarrow}(x,y=-\frac{L}{2})\equiv e^{-\frac{ B^{(2)}}{2\pi}(\frac{1}{x-i(-\frac{L}{2})})}F_{\downarrow}(x,-\frac{L}{2})=0$ 
 
\textbf{or}

$U_{\epsilon,\uparrow}(x,y=-\frac{L}{2})\equiv e^{-\frac{ B^{(2)}}{2\pi}(\frac{1}{x+i(-\frac{L}{2})})}F_{\epsilon\uparrow}(x,-\frac{L}{2})=  U_{\epsilon,\downarrow}(x,y=\frac{L}{2})\equiv e^{-\frac{ B^{(2)}}{2\pi}(\frac{1}{x-i\frac{L}{2}})}F_{\downarrow}(x,\frac{L}{2})=0$

Making the first choice, (both choices give the same eigenvalues and eigenfunction)  we compute the eigenfunctions $F_{\epsilon\uparrow}(x,y)$ and $F_{\epsilon\downarrow}(x,y)$ for $|y|>d$ using the boundary conditions :  
\begin{equation}  
F_{\epsilon\uparrow}(x,y=\frac{L}{2})=0;\hspace{0.4 in}
F_{\epsilon\downarrow}(x,y=-\frac{L}{2})=0 
\label{equabound}
\end{equation} 
Due to the fact that the solutions are restricted to $|y|>d$ no  conditions need to be imposed   at $x=y=0$.  
In the present case we consider a situation with a single dislocation. This is justified for a dilute concentration of  dislocations  typically separated by  a distance $l\approx 10^{-6} m$.  ( In principle we need at least two dislocations in order to satisfy the condition that the  sum of the Burger vectors is zero.) 
The eigenvalues are given by $\epsilon=\pm\hbar v_{F}\sqrt{p^2+q^2}$.  The value of $p$ is determined by the periodic boundary condition in the $x$ direction       $p(m)=\frac{2\pi}{L}m\equiv \frac{2\pi}{N a}m$, $m=0,1,...,(N-2),(N-1)$ and $a$ is the lattice constant  $a\approx \frac{2\pi}{\Lambda}$. The value of $q$ 
will be obtained  from the  vanishing boundary conditions  at $y=\pm\frac{L}{2}$.
The eigenfunctions $F_{\epsilon,\sigma}(x,y)$ will  be obtained using the  linear combination  of the  spinors  introduced in chapter $III$. In the Cartesian representation we can build four spinors  $\Gamma_{p,q}(x,y)$, $\Gamma_{p,-q}(x,y)$,$\Gamma_{-p,q}(x,y)$,$\Gamma_{-p,-q}(x,y)$  which are eigenstates of the chirality operator and are given by:
\begin{equation*}
\Gamma_{p,q}(x,y)=e^{i p x}e^{iq y} \left(\begin{array}{cc} 1\\ i e^{i\chi(p,q)}\end{array}
\right)
\end{equation*}
\begin{equation*}
\Gamma_{p,-q}(x,y)=e^{i p x}e^{-iq y}\left(\begin{array}{cc} 1\\ i e^{-i\chi(p,q)}\end{array}
\right)
\end{equation*}
\begin{equation*}
\Gamma_{-p,q}(x,y)=e^{i p x}e^{iq y}\left(\begin{array}{cc} 1\\ -i e^{-i\chi(p,q)}\end{array}
\right)
\end{equation*}
\begin{equation}
\Gamma_{-p,-q}(x,y)=e^{-i p x}e^{-iq y}\left(\begin{array}{cc} 1\\ -i e^{i\chi(p,q)}\end{array}
\right)
\label{spinors}
\end{equation}
where $tan[\chi(p,q)]=\frac{q}{p}$.

The $TI$  Hamiltonian  $h^{T.I}(x,y)=\hbar v_{F}[i\sigma^{1}\partial_{y} -i\sigma^{2}\partial_{x}]$ is invariant under the symmetry  the  operation  $x\rightarrow -x$ which is described by  the transformation $P_{x}$ :
$P_{x}x P^{-1}_{x}=-x$;  $P_{x}\sigma^{2} P^{-1}_{x}=-\sigma^{2}$; 
$P_{x}y P^{-1}_{x}=y$;  $P_{x}\sigma^{1} P^{-1}_{x}=\sigma^{1}$.

 The  edge Hamiltonian $h^{edge}$  contains in addition the term  $\sigma^{2}\delta(\vec{r})$ which changes sign  under the  symmetry  operation  $P_{x}$ . As a result  the symmetry operation  does not commute  with the edge Hamiltonian
$[h^{edge},P_{x}]\neq0$. 
This result demands  that  we  construct two independent   eigenfunctions $F^{(n=0,R)}_{p>0,q}(x,y)$ ($right-mover$) for $p>0$  and $ F^{(n=0,L)}_{-p>0,q}(x,y)$  ($left-mover$) $p<0$.
\begin{eqnarray}
&&F^{(n=0,R)}_{p>0,q}(x,y)=A(q)\Gamma_{p,q}(x,y)+B(q)\Gamma_{p,-q}(x,y)\nonumber\\&&
F^{(n=0,L)}_{-p>0,q}(x,y)=C(q)\Gamma_{-p,q}(x,y)+D(q)\Gamma_{-p,-q}(x,y)\nonumber\\&&
\end{eqnarray}
Employing  the boundary conditions given in equation $(29)$ we obtain the amplitudes $\frac{D(q)}{C(q)}$ ,  $\frac{B(q)}{A(q)}$  and the  discrete  momenta $q_{+}$.
Using  the pair $ \Gamma_{p,q}(x,y)$ , $\Gamma_{p,-q}(x,y)$ $p>0$ we obtain :
\begin{eqnarray}
&&F^{(n=0,R)}_{\epsilon(p>0,q_{+}),\uparrow}(x,y)=e^{ipx}e^{\frac{i}{2}\chi(p,q_{+})}[e^{i(q_{+}y-\frac{1}{2}\chi(p,q_{+}))}+(-1)^{k+1}e^{-i(q_{+}y-\frac{1}{2}\chi(p,q_{+}))}]; |y|>d\nonumber\\&&
F^{(n=0,R)}_{\epsilon(p>0,q_{+}),\downarrow}(x,y)=ie^{ipx}e^{\frac{i}{2}\chi(p,q_{+})}[e^{i(q_{+}y+\frac{1}{2}\chi(p,q_{+}))}+(-1)^{k+1}e^{-i(q_{+}y+\frac{1}{2}\chi(p,q_{+}))}]; |y|>d\nonumber\\&&
q\equiv q_{+}=\frac{\pi}{L}k+\frac{1}{L}\tan^{-1}(\frac{q_{+}}{p});k=1,2,3...;\tan[\chi(p,q_{+})]=(\frac{q_{+}}{p})\nonumber\\&&
\epsilon(p,q_{+})=\pm\hbar v_{F}\sqrt{(\frac{2\pi}{L}m)^2+q_{+}^2}\nonumber\\&&
\end{eqnarray}
Similarly, for the second  pair $ \Gamma_{-p,q}(x,y)$,$\Gamma_{-p,-q}(x,y)$, $p>0$  we obtain:
\begin{eqnarray}
&&F^{(n=0,L)}_{\epsilon(-p>0,q_{-}),\uparrow}(x,y)=e^{-ipx}e^{-\frac{i}{2}\chi(p,q_{-})}[e^{i(q_{-}y+\frac{1}{2}\chi(p,q_{-}))}+(-1)^{k+1}e^{-i(q_{-}y+\frac{1}{2}\chi(p,q_{-}))}]; |y|>d\nonumber\\&&
F^{(n=0,L)}_{\epsilon(-p>0,q_{-}),\downarrow}(x,y)=-ie^{-ipx}e^{-\frac{i}{2}\chi(p,q_{-})}[e^{i(q_{-}y-\frac{1}{2}\chi(p,q_{-}))}+(-1)^{k+1}e^{-i(q_{-}y-\frac{1}{2}\chi(p,q_{-}))}]; |y|>d\nonumber\\&&
q\equiv q_{-}=\frac{\pi}{L}k-\frac{1}{L}\tan^{-1}(\frac{q_{-}}{p});k=1,2,3...;\tan[\chi(p,q_{-})]=(\frac{q_{-}}{p})\nonumber\\&&
\epsilon(-p,q_{-})=\pm\hbar v_{F}\sqrt{(\frac{2\pi}{L}m)^2+q_{-}^2}\nonumber\\&&
\end{eqnarray}
For the state  with zero momentum $p=0$ we find:
\begin{eqnarray}
&&F^{(n=0,0)}_{\epsilon(p=0,q),\uparrow}(x,y)=2 e^{\frac{-i\pi}{4}}\cos[qy+\frac{\pi}{4}]; |y|>d\nonumber\\&&
F^{(n=0,0)}_{\epsilon(p=0,q),\downarrow}(x,y)=i2e^{\frac{-i\pi}{4}}\cos[qy-\frac{\pi}{4}]; |y|>d\nonumber\\&&
q=\frac{\pi}{2L}+\frac{\pi}{L}k;k=0,1,2,3...\nonumber\\&&
\epsilon(p=0,q)=\pm\hbar v_{F}|q|\nonumber\\&&
\end{eqnarray}
The eigenfunctions for the dislocation problem  for $|y|>d$   will be given in terms of the envelope functions $ e^{-\frac{B^{(2)}}{2\pi}(\frac{1}{x+iy})}$ , $e^{-\frac{B^{(2)}}{2\pi}(\frac{1}{x-iy})} $ ($U_{\epsilon,\uparrow}(x,y)=e^{-\frac{B^{(2)}}{2\pi}(\frac{1}{x+iy})}F_{\epsilon,\uparrow}(x,y)$, $U_{\epsilon,\downarrow}(x,y)=e^{-\frac{B^{(2)}}{2\pi}(\frac{1}{x-iy})}F_{\epsilon,\downarrow}(x,y)$).
 
The explicit solutions are given by : 

 $u^{(n=0,R)}_{\epsilon}(x,y)\equiv[U^{(n=0,R}_{\epsilon\uparrow}(x,y),U^{(n=0,R)}_{\epsilon\downarrow}(x,y)]^{T}$; $u^{(n=0,L)}_{\epsilon}(x,y)\equiv[U^{(n=0,L}_{\epsilon\uparrow}(x,y),U^{(n=0,L)}_{\epsilon\downarrow}(x,y)]^{T}$.

The components of the spinor are given by:
\begin{eqnarray}
&&U^{(n=0,R)}_{\uparrow}(x,y)\approx\frac{2 const.(B^{(2)})}{ G^{\frac{1}{4}}(x,y)L}
e^{-\frac{ B^{(2)}}{2\pi}(\frac{1}{x+iy})}F^{(n=0,R)}_{\epsilon(p>0,q_{+}),\uparrow}(x,y)\nonumber\\&&
U^{(n=0,R)}_{\downarrow}(x,y)\approx\frac{2 const.(B^{(2)})}{ G^{\frac{1}{4}}(x,y)L}
e^{-\frac{ B^{(2)}}{2\pi}(\frac{1}{x-iy})}F^{(n=0,R)}_{\epsilon(p>0,q_{+}),\downarrow}(x,y)\nonumber\\&&
U^{(n=0,L)}_{\uparrow}(x,y)\approx\frac{2 const.(B^{(2)})}{ G^{\frac{1}{4}}(x,y)L}
e^{-\frac{ B^{(2)}}{2\pi}(\frac{1}{x+iy})}F^{(n=0,L)}_{\epsilon(-p>0,q_{-}),\uparrow}(x,y)\nonumber\\&&
U^{(n=0,L)}_{\downarrow}(x,y)\approx\frac{2 const.(B^{(2)})}{ G^{\frac{1}{4}}(x,y)L}
e^{-\frac{ B^{(2)}}{2\pi}(\frac{1}{x-iy})}F^{(n=0,L)}_{\epsilon(-p>0,q_{-}),\downarrow}(x,y)\nonumber\\&&
U^{(n=0,0)}_{\uparrow}(x,y)\approx e^{-\frac{ B^{(2)}}{2\pi}(\frac{1}{x+iy})}F^{(n=0,0)}_{\epsilon(p=0,q),\uparrow}(x,y)\nonumber\\&&
U^{(n=0,0)}_{\downarrow}(x,y)\approx e^{-\frac{ B^{(2)}}{2\pi}(\frac{1}{x-iy})}F^{(n=0,0)}_{\epsilon(p=0,q),\downarrow}(x,y)\nonumber\\&&
\end{eqnarray}
where $G(x,y)=1-\frac{B^(2)}{2\pi}\frac{y}{\sqrt{2}(x^2+y^2)}$ is the Jacobian introduced by the edge dislocation.  The eigenstates are normalized and obey:$\int \,dx \int\,dy \sqrt{G(x,y)} (U^{(n=0,R)}_{\sigma}(x,y))^{*}  U^{(n=0,R)}_{\sigma'}(x,y)\approx \delta_{\sigma ,\sigma'}$, $\int \,dx \int\,dy \sqrt{G(x,y)} (U^{(n=0,L)}_{\sigma}(x,y))^{*}  U^{(n=0,L)}_{\sigma'}(x,y)\approx \delta_{\sigma ,\sigma'}$.
The  normalization  factor  $\frac{2 const.(B^{(2)} )}{L}\approx \frac{2  }{L}$,  has a weak  dependence  on the  Burger vector  $B^{(2)}$ . This dependence is a consequence of the Jacobian   $\sqrt{G}$ which affects  the normalization  constant. 

(The  multiplicative factor  $e^{-\frac{ B^{(2)}}{2\pi}(\frac{1}{x \pm iy})}$  gives rise to a weak non-orthogonality  between the states.
 This    non-orthogonality  of the linear independent  eigenfunctions  can be corrected  with the help of  the Grahm-Shmidt method.)
 
 For the present case,  backscattering is allowed but it is much weaker in comparison to regular metals. This is seen as follows:
Time reversal is not violated; due to the parity violation, the  eigenstates  $u^{(n=0,R)}_{\epsilon}(x,y)$ ,$u^{(n=0,L)}_{\epsilon}(x,y)$ are \textbf{not} related by a time reversal symmetry ($T u^{(n=0,R)}_{\epsilon}(x,y)\neq  u^{(n=0,L)}_{\epsilon}(x,y)$) . As a result, the backscattering potential $V_{p,-p}$ is controlled by a finite matrix element   between states with different eigenvalues  $\epsilon(-p,q_{-})\neq  \epsilon(p,q_{+})$ (contrary to regular metals  where the impurity potential   $V_{p,-p}$ connects states with the same energy). In the present case $|\epsilon(-p,q_{-}) -   \epsilon(p,q_{+})|=\hbar v_{F}|[\sqrt{(\frac{2\pi}{L}m)^2+q_{-}^2}-\sqrt{(\frac{2\pi}{L}m)^2+q_{+}^2}]|\neq0$  the eigenvalues are not equal, therefore the finite matrix element controlled by the   backscattering potential  $V_{p,-p}$ gives rise only  to   a second order backscattering effect!

\vspace{0.1 in}
  
\textbf{IIIE- The  circular  contours-the wave function for  $n\neq0$}

\vspace{0.1 in}

 The equation $I(z,\overline{z})=e^{2\frac{B^{(2)}}{\pi}(\frac{ iy(s)}{x^2(s)+y^2(s)})}  =e^{i 2\pi n}$  gives the set  of ring  contours  for $n=\pm1,\pm2,\pm3,...$ shown in figure 1.
The radius  $R_{g}$ for the fundamental contour($n=1$) is represented  in terms of the  Burger vector  $B^{(2)}$,   $R_{g}=\frac{B^{(2)}}{2\pi^2}$ and $R_{g}(n)=\frac{R_{g}}{|n|}$.
\begin{equation}
(x(s))^2+ (y(s)\pm R_{g}(n))^2 =(R_{g}(n))^2
\label{contour}
\end{equation}
The centers of the contours are given by :$[\bar{x},\bar{y}]=[0,R_{g}(n)]$  for $n\neq 0$. When $n>0$ the center of the contours has positive coordinates (upper contour) and for $n<0$  the center has negative coordinates (lower contour). 
Each contour is characterized by a circle with a radius $R_{g}(n)\equiv\frac{R_{g}}{|n|}$  centered at  $[\bar{x}=0,\bar{y}=R_{g}(n)]$. The contour is parametrized in terms of  the  arc length $0\leq s< 2\pi \frac{R_{g}}{|n|}$ which is  equivalent to  $0\leq\varphi<2\pi$ .
Each contour is parametrized  by  $\vec{r}(s)\equiv[x(s),R_{g}(n)+y(s)]$ where $x(s)= R_{g}(n) \cos[\frac{s}{R_{g}(n)}]\equiv\ R_{g}(n)\cos[\varphi]$ and $y(s)=  R_{g}(n) \sin[\frac{s}{R_{g}(n)}]\equiv\ R_{g}(n)\sin[\varphi]$. We will extend this curve to a two dimensional strip with the  coordinate $u$ in the normal direction:
For the curve  curve $\vec{r}(s)=[x(s),y(s)]$ we will use the tangent  $\vec{t}(s)$  and the normal vector  $\vec{N}(s)$  
 Therefore,   the  two dimensional region in the vicinity of the one parameter curve $\vec{r}(s)$ is replaced by   $\vec{r}(s)\rightarrow  \vec{R}(s,u)=\vec{r}(s)+u \vec{N}(s)$. 
\begin{eqnarray}
&&x(s,u)=R_{g}(n) \cos[\frac{s}{R_{g}(n)}]+u\cos[\frac{s}{R_{g}(n)}]\nonumber\\&&
y(s,u)=R_{g}(n) \sin[\frac{s}{R_{g}(n)}]+u\sin[\frac{s}{R_{g}(n)}]\nonumber\\&&
\end{eqnarray}
We will restrict the width $|u|$ such that $e^{i 2\pi n}e^{ \pm i\eta}\approx 1 $  where $\eta$ obeys  $\eta <\frac{\pi}{4}<1$ ,
  $|u|\leq \frac{R_{g}(n)}{1-\frac {\eta}{2\pi n}}-R_{g}(n)\approx R_{g}(n)(\frac {\eta}{2\pi n})< \frac{R_{g}(n)}{8n}$.  
In these new coordinates, the Dirac equation is approximated for  $|u|\leq R_{g}(n)(\frac {\eta}{2\pi n})=\frac{D(n)}{2}$ by :
\begin{eqnarray}
&&\epsilon F_{\epsilon \uparrow}(s,u)=-I(s,u)e^{-i\frac{s}{R_{g}(n)}}[\partial_{u}-\frac{i}{1+\frac{u}{R_{g}(n)}}\partial_{s}]F_{\epsilon\downarrow}(s,u)\approx -e^{-i\frac{s}{R_{g}(n)}}[\partial_{u}-i\partial_{s}]F_{\epsilon\downarrow}(s,u) \nonumber\\&&
\epsilon F_{\epsilon\downarrow}(s,u)=(I(s,u))^{*}e^{i\frac{s}{R_{g}(n)}}[\partial_{u}+\frac{i}{1+\frac{u}{R_{g}(n)}}\partial_{s}]F_{\epsilon\uparrow}(s,u)\approx e^{i\frac{s}{R_{g}(n)}}[\partial_{u}+i\partial_{s}]F_{\epsilon\uparrow}(s,u)\nonumber\\&&
\end{eqnarray}

\vspace{0.1 in}

\textbf{The solution  for the contour  $n\neq0$, $0\leq s <2\pi R_{g}(n)$; $|u|\leq \frac{D(n)}{2}$}

\vspace{0.1 in}

The periodicity in $s$ allows us to  represent the  eigenfunctions in  the form: $F_{\epsilon\uparrow}(s,u)=\sum_{j=-\infty}^{\infty}\sum_{q}e^{i j(\frac{s}{R_{g}(n)})}e^{iqu}F_{\epsilon\uparrow}(j,q)$
and $F_{\epsilon\downarrow}(s,u)=\sum_{j=-\infty}^{\infty}\sum_{q}e^{i (j+1)(\frac{s}{R_{g}(n)})}e^{iqu}F_{\epsilon\downarrow}(j,q)$. We find:
\begin{eqnarray}
&&\epsilon F_{\uparrow}(\epsilon;j,q)=(i q +\frac{j}{R_{g}(n)})F_{\downarrow}(\epsilon;j,q)\nonumber\\&&
\epsilon F_{\downarrow}(\epsilon;j,q)=(i q +\frac{j+1}{R_{g}(n)})F_{\uparrow}(\epsilon;j,q)\nonumber\\&&
\end{eqnarray}
The determinant of  the  two  equations  determines the relation between the eigenvalue $\epsilon$, the transverse momentum $Q(\epsilon)$ and the eigenfunctions $F_{\epsilon\downarrow}(j,q)$,$F_{\epsilon\uparrow}(j,q)$.  The eigenvalues are degenerate    and obey : $\epsilon(j=l;k)=\epsilon(j=-(l+1);k)$ ,where $l\geq0$. 
\begin{eqnarray}
&&q\equiv\frac{-i}{2 R_{g}(n)}\pm Q(\epsilon);\hspace{0.2 in}Q(\epsilon)=\sqrt{\epsilon^{2}-(\frac{l+\frac{1}{2}}{ R_{g}(n)})^2}\nonumber\\&&
F_{\epsilon}(l,q)\equiv[F_{\epsilon\uparrow}(l,q),F_{\epsilon\downarrow}(l,q)]^{T}\propto [1,e^{-i\kappa(Q,l)}]^{T};\hspace{0.1 in}  \kappa(Q,l)=tan^{-1}(\frac{Q R_{g}(n)}{l+\frac{1}{2}})\nonumber\\&& 
\end{eqnarray}
The value of the transversal momentum $ Q(\epsilon) $ will be determined  from the  boundary conditions at $\pm\frac{D(n)}{2}$. 
We will introduce a polar angle  $\theta$ measured with respect the Cartesian axes:
The  angle $0<\varphi(n=1)\leq 2\pi$  for  the upper contour  $n=1$  centered  at $[\overline{x}=0,\overline{y}=R_{g}]$ is described by the polar coordinate $ 0<\theta  \leq \pi$ measured from the center of the Cartesian coordinate $[0,0]$. The lower contour centered at $[\overline{x}=0,\overline{y}=-R_{g}]$  characterized by  the angle  $0<\varphi(n=-1)\leq 2\pi$  is described  by the polar angle $\theta$ restricted  to  $\pi<\theta \leq2\pi$. We establish the correspondence between  $\varphi(n=\pm1)$ and $\theta$:
\begin{eqnarray}
&&\varphi(n=1)=2\theta+\frac{3\pi}{2} \hspace{0.1 in}  for  \hspace{0.1 in}the \hspace{0.1 in} upper \hspace{0.1 in} contour \hspace{0.1 in} n=1, \hspace{0.1 in}  0<\theta  \leq \pi \nonumber\\&&
\varphi(n=-1)=2\theta+\frac{3\pi}{2}+\pi \hspace{0.1 in} for  \hspace{0.1 in} the \hspace{0.1 in} lower \hspace{0.1 in} contour \hspace{0.1 in} n=-1 , \hspace{0.1 in}0<\theta \leq\pi \nonumber\\&&
\end{eqnarray}
Following the discussion from the  previous chapter we will introduce the following boundary conditions:
\begin{eqnarray}  
&&F^{(n=1)}_{\epsilon\uparrow}(s,u=\frac{D}{2})=0;\hspace{0.4 in}
F^{(n=1)}_{\epsilon\downarrow}(s,y=-\frac{D}{2})=0 \nonumber\\&&
F^{(n=-1)}_{\epsilon\uparrow}(s,u=-\frac{D}{2})=0;\hspace{0.4 in}
F^{(n=-1)}_{\epsilon\downarrow}(s,y=\frac{D}{2})=0 \nonumber\\&&
D(n=\pm1)\equiv D
\end{eqnarray}
For the two contours $n=\pm1$ we  introduce eight spinors  
$\Gamma^{(n=\pm1)}_{l,Q}(\varphi(n=\pm1),u)$,$\Gamma^{(n=\pm1)}_{l,-Q}(\varphi(n=\pm1),u)$, $\Gamma^{(n=\pm1)}_{-l,Q}(\varphi(n=\pm1),u) $,  $\Gamma^{(n=\pm1)}_{-l,-Q}(\varphi(n=\pm1),u)$. Using this spinor  we will compute    the eigenfunctions.  For the case $n=0$ we had only four spinors   given in equation $(30)$. The four spinors have been used to construct the eigenfunctions $F^{(n=0,R)}_{p>0,q}(x,y)$  for $p>0$  and $ F^{(n=0,L)}_{-p>0,q}(x,y)$ . Due to the fact that for each $n\neq 0$ we have two contours $n=\pm$ we have eight spinors  which will be used to construct the eigenfunctions.

\begin{equation*}
\Gamma^{(n=\pm1)}_{l,Q}(\varphi(n=\pm1),u)=e^{i l (\varphi(n=\pm1))}e^{iQ u} \left(\begin{array}{cc} 1\\ e^{i (\varphi(n=\pm1))} e^{-i\kappa(l,Q)}\end{array}
\right)
\end{equation*}
\begin{equation*}
\Gamma^{(n=\pm1)}_{l,-Q}(\varphi(n=\pm1),u)=e^{i l(\varphi(n=\pm1))}e^{-iQ u} \left(\begin{array}{cc} 1\\ e^{i (\varphi(n=\pm1))} e^{i\kappa(l,Q)}\end{array}
\right)
\end{equation*}
\begin{equation*}
\Gamma^{(n=\pm1)}_{-l,Q}(\varphi(n=\pm1),u)=e^{-i l(\varphi(n=\pm1))}e^{iQ u} \left(\begin{array}{cc} 1\\- e^{-i (\varphi(n=\pm1))} e^{i\kappa(l,Q)}\end{array}
\right)
\end{equation*}
\begin{equation}
\Gamma^{(n=\pm1)}_{-l,-Q}(\varphi(n=\pm1),u)=e^{-i l (\varphi(n=\pm1))}e^{-iQ u} \left(\begin{array}{cc} 1\\- e^{-i (\varphi(n=\pm1))} e^{-i\kappa(l,Q)}\end{array}
\right)
\label{spinors}
\end{equation}
Using the vanishing boundary condition given in equation $(42)$ we construct for this case similar spinors as the one given in equation $(31)$.  In the present case  we have for each $n\neq 0$ two contours, therefore the number of spinors will be doubled. We find  instead of the eigenfunction given in equation $(33)$   two sets of eigenfunctions with  momentum $Q_{-}$ (which replaces $q_{-}$ , see $(33)$)  and $Q_{+}$ (which replaces  $q_{+}$ , see $(32)$) .

Using the boundary conditions given in eq.$(35)$ we determine the quantization conditions $Q_{-}$,$Q_{+}$  and the eigenfunctions for  the $ n=1$ and $n=-1$ contours.  
\begin{eqnarray}
&&Q_{-}=\frac{\pi}{D}k-\frac{1}{D}\tan^{-1}(\frac{Q_{-}R_{g}(1)}{l+\frac{1}{2}}),k=1,2,3...;
\tan[\kappa(l,Q_{-})]=(\frac{Q_{-}R_{g}(1)}{l+\frac{1}{2}})\nonumber\\&&
\epsilon(l,Q_{-})=\pm\hbar v_{F}\sqrt{(\frac{l+\frac{1}{2}}{ R_{g}(1)})^2+Q_{-}^2}\nonumber\\&&
Q_{+}=\frac{\pi}{D}k+\frac{1}{D}\tan^{-1}(\frac{Q_{+}R_{g}(1)}{l+\frac{1}{2}}), k=1,2,3...\nonumber\\&&  \tan[\kappa(l,Q_{+})]=(\frac{Q_{+}R_{g}(1)}{l+\frac{1}{2}})\nonumber\\&&
\epsilon(l,Q_{+})=\pm\hbar v_{F}\sqrt{(\frac{l+\frac{1}{2}}{ R_{g}(n)})^2+Q_{+}^2}
\end{eqnarray}
 
Using the fact that the combined wave function on the contours  $n=1$ and $n=\pm1$ must be finite we obtain two sets of wave functions. We include the envelope function and obtain the wave function for $ Q_{-}$ and $Q_{+}$:
The envelope functions $e^{-\frac{ B^{(2)}}{2\pi}(\frac{1}{x+iy})}$ , $e^{-\frac{ B^{(2)}}{2\pi}(\frac{1}{x-iy})}$  when projected to the contours  take a complicated form. The envelope functions  can be expressed  in terms of the  functions  $\eta(u)$ and $\zeta(\theta,u)$:
\begin{eqnarray} 
&&\eta(u)=\frac{R_{g}(1)}{R_{g}(1)+u},\frac{|u|}{R_{g}(1)}<1\nonumber\\&& \zeta(\theta,u)=\frac{-B^{(2)}}{2\pi(R_{g}(1)+u)((\sin[2\theta])^2+(\eta(u)-\cos[2\theta])^2)}\nonumber\\&&
\end{eqnarray}
We find for $Q_{-}$:
\begin{eqnarray}
&&U_{\epsilon(l,Q_{-})\uparrow}(\theta,u)= G^{\frac{-1}{4}}(\theta,u)\cdot[U^{(even,k)}_{\epsilon(l,Q_{-})\uparrow}(\theta,u)+
U^{(odd,k)}_{\epsilon(l,Q_{-})\uparrow}(\theta,u)];\nonumber\\&&
U^{(even,k)}_{\epsilon(l,Q_{-})\uparrow}(\theta,u)=
2i e^{\frac{-i}{2} \kappa(l,Q_{-})}[e^{\zeta(\theta,u)\sin[2\theta]}e^{-i\zeta(\theta,u)(\eta(u)-\cos[2\theta])}e^{il(2\theta+\frac{3\pi}{2})}\sin[Q_{-}u+\frac{1}{2}\kappa(l,Q_{-})]\nonumber\\&&
+(-1)^l e^{-\zeta(\theta,u)\sin[2\theta]}e^{-i\zeta(\theta,u)(-\eta(u)+\cos[2\theta])}e^{-il(2\theta+\frac{3\pi}{2})}\sin[Q_{-}u-\frac{1}{2}\kappa(l,Q_{-})]];\nonumber\\&&
U^{(odd,k)}_{\epsilon(l,Q_{-})\uparrow}(\theta,u)=
2 e^{\frac{-i}{2} \kappa(l,Q_{-})}[e^{\zeta(\theta,u)\sin[2\theta]}e^{-i\zeta(\theta,u)(\eta(u)-\cos[2\theta])}e^{il(2\theta+\frac{3\pi}{2})}\cos[Q_{-}u+\frac{1}{2}\kappa(l,Q_{-})]\nonumber\\&&
+(-1)^l e^{-\zeta(\theta,u)\sin[2\theta]}e^{-i\zeta(\theta,u)(-\eta(u)+\cos[2\theta])}e^{-il(2\theta+\frac{3\pi}{2})}\cos[Q_{-}u-\frac{1}{2}\kappa(l,Q_{-})]];\nonumber\\&&
U_{\epsilon(l,Q_{-})\downarrow}(\theta,u)=G^{\frac{-1}{4}}(\theta,u)\cdot[U^{(even,k)}_{\epsilon(l,Q_{-})\downarrow}(\theta,u)+
U^{(odd,k)}_{\epsilon(l,Q_{-})\uparrow}(\theta,u)];\nonumber\\&&
U^{(even,k)})_{\epsilon(l,Q_{-})\downarrow}(\theta,u)=
2i e^{\frac{-i}{2} \kappa(l,Q_{-})}[e^{\zeta(\theta,u)\sin[2\theta]}e^{i\zeta(\theta,u)(\eta(u)-\cos[2\theta])}e^{i((l+1)(2\theta+\frac{3\pi}{2}))}\sin[Q_{-}u-\frac{1}{2}\kappa(l,Q_{-})]\nonumber\\&&
-(-1)^l e^{-\zeta(\theta,u)\sin[2\theta]}e^{i\zeta(\theta,u)(-\eta(u)+\cos[2\theta])}e^{-i((l+1)(2\theta+\frac{3\pi}{2}))}\sin[Q_{-}u+\frac{1}{2}\kappa(l,Q_{-})]];\nonumber\\&&
U^{(odd,k)}_{\epsilon(l,Q_{-})\downarrow}(\theta,u)=2e^{\frac{-i}{2} \kappa(l,Q_{-})}[e^{\zeta(\theta,u)\sin[2\theta]}e^{i\zeta(\theta,u)(\eta(u)-\cos[2\theta])}e^{i((l+1)(2\theta+\frac{3\pi}{2}))}\cos[Q_{-}u-\frac{1}{2}\kappa(l,Q_{-})]\nonumber\\&&
-(-1)^l e^{-\zeta(\theta,u)\sin[2\theta]}e^{i\zeta(\theta,u)(-\eta(u)+\cos[2\theta])}e^{-i((l+1)(2\theta+\frac{3\pi}{2}))}\cos[Q_{-}u+\frac{1}{2}\kappa(l,Q_{-})]];\nonumber\\&&
\end{eqnarray} 
Similarly for  $Q_{+}$we obtain the wave function:
\begin{eqnarray}
&&U_{\epsilon(l,Q_{+})\uparrow}(\theta,u)= G^{\frac{-1}{4}}(\theta,u)\cdot[U^{(even,k)}_{\epsilon(l,Q_{+})\uparrow}(\theta,u)+
U^{(odd,k)}_{\epsilon(l,Q_{+})\uparrow}(\theta,u)];\nonumber\\&&
U^{(even,k)}_{\epsilon(l,Q_{+})\uparrow}(\theta,u)=
2i e^{\frac{-i}{2} \kappa(l,Q_{+})}[(-1)^{l}e^{-\zeta(\theta,u)\sin[2\theta]}e^{-i\zeta(\theta,u)(-\eta(u)+\cos[2\theta])}e^{il(2\theta+\frac{3\pi}{2})}\sin[Q_{+}u+\frac{1}{2}\kappa(l,Q_{+})]\nonumber\\&&
+ e^{\zeta(\theta,u)\sin[2\theta]}e^{-i\zeta(\theta,u)(\eta(u)-\cos[2\theta])}e^{-il(2\theta+\frac{3\pi}{2})}\sin[Q_{+}u-\frac{1}{2}\kappa(l,Q_{+})]];\nonumber\\&&
U^{(odd,k)}_{\epsilon(l,Q_{+})\uparrow}(\theta,u)=
2  e^{\frac{-i}{2} \kappa(l,Q_{+})}[(-1)^{l}e^{-\zeta(\theta,u)\sin[2\theta]}e^{-i\zeta(\theta,u)(-\eta(u)+\cos[2\theta])}e^{il(2\theta+\frac{3\pi}{2})}\cos[Q_{+}u+\frac{1}{2}\kappa(l,Q_{+})]\nonumber\\&&
+ e^{\zeta(\theta,u)\sin[2\theta]}e^{-i\zeta(\theta,u)(\eta(u)-\cos[2\theta])}e^{-il(2\theta+\frac{3\pi}{2})}\cos[Q_{+}u-\frac{1}{2}\kappa(l,Q_{+})]]; \nonumber\\&&    
U_{\epsilon(l,Q_{+})\downarrow}(\theta,u)= G^{\frac{-1}{4}}(\theta,u)\cdot[U^{(even,k)}_{\epsilon(l,Q_{+})\downarrow}(\theta,u)+
U^{(odd,k)}_{\epsilon(l,Q_{+})\downarrow}(\theta,u)];\nonumber\\&&
U^{(even,k)}_{\epsilon(l,Q_{+})\downarrow}(\theta,u)=
2i e^{\frac{-i}{2} \kappa(l,Q_{+})}[-(-1)^{l}e^{-\zeta(\theta,u)\sin[2\theta]}e^{i\zeta(\theta,u)(-\eta(u)+\cos[2\theta])}e^{i(l+1)(2\theta+\frac{3\pi}{2})}\sin[Q_{+}u-\frac{1}{2}\kappa(l,Q_{+})]\nonumber\\&&
+ e^{\zeta(\theta,u)\sin[2\theta]}e^{i\zeta(\theta,u)(\eta(u)-\cos[2\theta])}e^{-i(l+1)(2\theta+\frac{3\pi}{2})}\sin[Q_{+}u+\frac{1}{2}\kappa(l,Q_{+})]];\nonumber\\&&
U^{(odd,k)}_{\epsilon(l,Q_{+})\downarrow}(\theta,u)=
2 e^{\frac{-i}{2} \kappa(l,Q_{+})}[-(-1)^{l}e^{-\zeta(\theta,u)\sin[2\theta]}e^{i\zeta(\theta,u)(-\eta(u)+\cos[2\theta])}e^{i(l+1)(2\theta+\frac{3\pi}{2})}\cos[Q_{+}u-\frac{1}{2}\kappa(l,Q_{+})]\nonumber\\&&
+ e^{\zeta(\theta,u)\sin[2\theta]}e^{i\zeta(\theta,u)(\eta(u)-\cos[2\theta])}e^{-i(l+1)(2\theta+\frac{3\pi}{2})}\cos[Q_{+}u+\frac{1}{2}\kappa(l,Q_{+})]];\nonumber\\&&    
\end{eqnarray}
where $G^{\frac{-1}{4}}(\theta,u)$  is  the Jacobian transformation induced by the metric tensor.

\vspace{0.2 in}

\textbf{IV -Computation of the  STM  density of states}

\vspace{0.2 in}

\textbf{A-Description of the STM method}

\vspace{0.1 in}

The STM tunneling current $I$ is a function of the bias voltage $V$ which  gives spatial and spectroscopic information about the  electronic surface states. At zero temperature, the derivative of the current with respect the bias voltage  $V$ is given in term of the single particles eigenvalues: $\epsilon(m,q_{-})=\pm\hbar v_{F}\sqrt{(\frac{2\pi}{L}m)^2+q_{-}^2}$,   $\epsilon(m,q_{+})=\pm\hbar v_{F}\sqrt{(\frac{2\pi}{L}m)^2+q_{-}^2}$ ,$m=0,1,2,3...$  for    contour $n=0$.
For the upper and lower circular contours $n=\pm1$,  we have :$\epsilon(l,Q_{-})=\pm\hbar v_{F}\sqrt{(\frac{l+\frac{1}{2}}{ R_{g}(1)})^2+Q_{-}^2}$ ,$\epsilon(l,Q_{+})=\pm\hbar v_{F}\sqrt{(\frac{l+\frac{1}{2}}{ R_{g}(1)})^2+Q_{+}^2}$  ,$l=0,1,2,3..$.
The $STM$ density of states  is computed for a voltage $V$ between the $STM$ tip and the sample. The  tunneling current is   a function of the   bias voltage  $V$  and the  chemical potential  $\mu>0$  \cite{kittel}: 
\begin{eqnarray}
&&\frac{d I}{dV}\propto D(E=eV;s,u)\equiv \sum_{n} D^{(n)}(E=eV;s,u)= \nonumber\\&&= \sum_{\eta=\pm}[\sum_{m}\sum_{q_{r}=q_{+},q_{-}}\sum_{\sigma}|U^{(n=0;m,q_{r})}_{\sigma}(x,y)|^{2}\delta[eV +\mu -\eta\hbar v_{F}\sqrt{(\frac{2\pi}{L}m)^2+q_{r}^2}]\nonumber\\&&+\sum_{n=\pm1}\sum_{l}\sum_{Q_{r}=Q_{+},Q_{-}}\sum_{\sigma}|U^{(n=\pm1;l,Q_{r})}_{\sigma}(\theta,u)|^{2}\delta[eV +\mu -\eta\hbar v_{F}\sqrt{(\frac{l+\frac{1}{2}}{ R_{g}(1)})^2+Q_{r}^2}]]\nonumber\\&&
\end{eqnarray}
($\eta=+$ corresponds to electrons  with energy $0<\epsilon\leq \mu$ and  $\eta=-$ corresponds to electrons below the Dirac point $\epsilon<0$.  For the  rest   of this   paper  we will take the chemical potentials to be  $\mu= 120mV$  (this is typical value for the $TI$ ). We will neglect   the states with $\eta=-$ which correspond to particles below the Dirac cone.
The   density of states at the  tunneling energy $eV$  is weighted  by the probability density of  the $STM$ tip  at position $[x,y]$  for n=0. The contours for  $n=\pm1$    will be parametrized  in terms of  the polar angle  $\theta$ and transverse coordinate $u$.
The proportionality factor $J$ for  the tunneling probability (not shown  in the equation )  $\frac{d I}{dV}=J  D(V;x,y)$  is a function of the distance between the tip and the sample.  The notation   $D^{(n)}(V;x,y)$ represents  the tunneling  density for the different contours.




\vspace{0.1 in}

\textbf{IVB-The  tunneling density of states $D^{(n=0)}(V;x,y)$  for   $n=0$}

\vspace{0.1 in}

Summing up the single particle states  weighted with occupation  probability  $|U^{(n=0;m,q_{r})}_{\sigma}(x,y)|^{2}$, we  obtain a space dependent  density of states for the  two dimensional boundary surface ,$\frac{-L}{2}\leq x\leq\frac{L}{2}$ and the coordinate $y$ is restricted to the regions $\frac{d}{2}<y\leq \frac{L}{2}$  and $\frac{-L}{2}<y\leq \frac{d}{2}$.  
We will perform the computation at the thermodynamic limit, namely we replace the discrete momentum $\frac{\pi}{L}k$ by $Y=\frac{k}{N}$  and $\frac{2\pi}{L}m$ by $X= \frac{m}{N}$ where $N=\frac{L}{a}$. We find for  the dimensionless momentum $\hat{q}\equiv q a$  the equations :
$ \hat{q}_{\pm}(Y)=\pi Y \pm\frac{1}{N}\tan^{-1}[\frac{\hat{q}_{\pm}(Y)}{2\pi X}]$ where   $2\pi X=pa=\hat{p}$.  As a result we obtain the following density of states $\frac{\partial \hat{q}_{\pm}}{\partial Y}$
\begin{eqnarray} 
&&[\frac{\partial \hat{q}_{+}}{\partial Y}]^{-1}=\frac{1}{\pi}\frac{\hat{q}_{+}^{2}+\hat{p}^{2}-\frac{1}{N}\hat{p}}{\hat{q}_{+}^{2}+\hat{p}^{2}}\nonumber\\&&
[\frac{\partial \hat{q}_{-}}{\partial Y}]^{-1}=\frac{1}{\pi}\frac{\hat{q}_{-}^{2}+\hat{p}^{2}+\frac{1}{N}\hat{p}}{\hat{q}_{-}^{2}+\hat{p}^{2}}\nonumber\\&&
\end{eqnarray}
Using this results, we  compute  the tunneling density of states in terms of the  energy  $\mu+eV$  measured with respect the chemical potential $\mu$ and  the  transverse energy $\epsilon_{\bot}\equiv \hbar v_{F}q_{\pm}$.
\begin{eqnarray}
&&D^{(n=0)}(V;x,y)= (\frac{L}{h v_{F}})^{2}(\frac{B^{(2)}}{L})^2   \frac{1}{4\sqrt{G(x,y)}}e^{\frac{-B^{(2)}}{\pi}(\frac{x}{x^2+y^2+a^2})}[\int_{0}^{E_{max.}}\,d \epsilon_{\bot}\frac{(\mu+eV)}{\sqrt{(\mu+eV)^2-\epsilon_{\bot}^2}}\dot\nonumber\\&&[\frac{1}{2}(1+\frac{1}{\pi}\frac{hv_{F}}{L (\mu+V)}\sqrt{1-(\frac{\epsilon_{\bot} }{\mu+V})^2}\hspace{0.1 in})+ \frac{1}{2}(1-\frac{1}{\pi}\frac{hv_{F}}{L (\mu+V)}\sqrt{1-(\frac{\epsilon_{\bot} }{\mu+V})^2}\hspace{0.1 in})] \nonumber\\&&+
\frac{h v_{F}}{L}(H[\mu+V-\frac{h v_{F}}{2 L}]-H[\mu+eV-E_{max}])\cdot((\cos[\frac{(\mu+eV)}{\hbar v_{F}}y-\frac{\pi}{4}])^2+(\cos[\frac{(\mu+eV)}{\hbar v_{F}}y-\frac{\pi}{4}])^{2})]\nonumber\\&&=
(\frac{L}{h v_{F}})^{2}(\frac{B^{(2)}}{L})^2   \frac{1}{4\sqrt{G(x,y)}}e^{\frac{-B^{(2)}}{\pi}(\frac{x}{x^2+y^2+a^2})}[\int_{0}^{E_{max.}}\,d \epsilon_{\bot}\frac{(\mu+eV)}{\sqrt{(\mu+eV)^2-\epsilon_{\bot}^2}} \nonumber\\&&+
\frac{h v_{F}}{L}(H[\mu+V-\frac{h v_{F}}{2 L}]-H[\mu+eV-E_{max}])\cdot((\cos[\frac{(\mu+eV)}{\hbar v_{F}}y-\frac{\pi}{4}])^2+(\cos[\frac{(\mu+eV)}{\hbar v_{F}}y-\frac{\pi}{4}])^{2})]=\nonumber\\&&
(\frac{L}{h v_{F}})^{2}(\frac{B^{(2)}}{L})^2   \frac{1}{4\sqrt{G(x,y)}}e^{\frac{-B^{(2)}}{\pi}(\frac{x}{x^2+y^2+a^2})}[\frac{\pi}{2}(\mu+eV)+\frac{h v_{F}}{L}(H[\mu+V-\frac{h v_{F}}{2 L}]-H[\mu+eV-E_{max}])]\nonumber\\&&for\hspace{0.2 in}  |y|>d
\nonumber\\&&
\end{eqnarray}
$H[\mu+eV-\frac{h v_{F}}{2 L}]$ is the step function  which is one for  $\mu+eV-\frac{h v_{F}}{2 L}\geq 0$ and  zero otherwise. $a=\frac{2\pi}{\Lambda}$ is the short distance cut-off and $E_{max}=\hbar v_{F}\Lambda<0.3 eV$ is the maximal energy  which restricts the validity of the Dirac model.  We observe  in the second  line that the  asymmetry in the density of states  $1\pm\frac{1}{\pi}\frac{hv_{F}}{L (\mu+V)}\sqrt{1-(\frac{\epsilon_{\bot} }{\mu+V})^2}\hspace{0.1 in})$ cancels.  

Equation $(51)$   shows  that the tunneling density of states is linear in the energy   $\mu+eV$  (in the present case we have looked only for energies above the Dirac cone ). For the   chemical potential $\mu=120mV$,  the zero energy corresponds to the Voltage $V=-120 mV$. The tunneling density of states has a constant part at energies $\frac{h v_{F}}{2 L}\approx 0.2 mV$  for $-120 mV<V<-119.8mV$. For $V>-119.8mV$ the density of states is proportional to $\mu+eV$.

In figure $2$ we have plotted the tunneling density of states as a function of the coordinates $x$ and $y$. The shape of the plot is governed by the the multiplicative  factor $e^{-\frac{B^{(2)}}{\pi}(\frac{x}{x\pm i y})}$ which governs the solutions in eq.$(35)$. We observe that the density of state is  maximal in the region $|y|<10 B^{(2)} $. 

Figure $3$ shows the dependence on the voltage $V$ and coordinate $y$. We observe the linear increase in the tunneling density of states which is maximal  in the region $|y|<10 B^{(2)}$.

\vspace{0.1 in} 

\textbf{IVC-The tunneling  density of states $D^{(n=0)}(V,x,y;\vec{r}_{1},..\vec{r}_{2M})$  for $2M$ dislocations.}

\vspace{0.1 in}

For many dislocations which  satisfy  $\sum_{w=1}^{2M}B^{(2,w)}=0$  ( sum of the Burger vectors is zero ) with  the core  centered at $[x_{w},y_{w}]$ ,$w=1,2..2M$ the coordinate $\vec{r}=(x,y)\rightarrow [X(\vec{r}),Y(\vec{r})]$ is replaced by  $[X(\vec{r})=x,Y(\vec{r})=y+\sum_{w}\frac{B^{(2,w)}}{2\pi}\tan^{-1}(\frac{y-y_{w}}{x-x_{w}})]$.
Following the method used previously,  we find the edge Hamiltonian with many dislocations  takes the form: 
\begin{equation}
h^{edge}(w=1,2...2M)\approx i\sigma^{1}[\partial_{y} -\frac{i}{2} \sum_{w=1}^{2M}\sigma^{3}B^{(2,w)}\delta^{2}(\vec{r}-\vec{r}_{w})] -i\sigma^{2}\partial_{x}
\label{Burgermany}
\end{equation}
As a result, the wave functions are given by: 
\begin{eqnarray}
&&U^{(n=0,w=1,2...2M)}_{\uparrow}(x,y)\propto
\prod_{w=1,2...2M}e^{-\frac{ B^{(2)}}{2\pi}(\frac{1}{(x-x_{w})+i(y-y_{w})})}F^{(n=0)}_{\uparrow}(x,y)\nonumber\\&&
U^{(n=0,w=1,2...2M)}_{\downarrow}(x,y)\propto
\prod_{w=1,2...2M}e^{-\frac{ B^{(2)}}{2\pi}(\frac{1}{(x-x_{w})-i(y-y_{w})})}F^{(n=0)}_{\downarrow}(x,y)
\end{eqnarray}
Using these wave functions, we find that the tunneling density of states is given by:
\begin{equation}
D^{(n=0)}(V,x,y;\vec{r}_{1},..\vec{r}_{2M})\propto \prod_{w=1,2...2M}e^{-\frac{ B^{(2)}}{\pi}(\frac{(x-x_{w})}{(x-x_{w})^2+(y-y_{w})^2+a^2})}
\label{densityw}
\end{equation}
In figure $4$ we show the tunneling density of states for an even number of dislocations in the $y$ directions which have the core on the $y=0$ axes ($\vec{r}_{w}=[x_{w},y_{w}=0]$, $w=1,2,3,...2M$). When the coordinate of the  $w=1,2,3,...2M$  dislocations  is replaced by a continuum  variable $w$ which  can  be described    by a domain  a wall model:
$h^{domain-wall}(x,y)= \hbar v_{F}[-i\sigma^{1}\partial_{y} +i\sigma^{2}\partial_{x}- \sigma^{3}\kappa M(y)]$ where  $M(y)=sgn[y]|M(y)|$   \cite{Jackiw}.

Using this model find that the tunneling density of states density  $D^{domain-wall}(V;x,y)$     confined to   $|y|< W$ (the width $W$ depends on the explicit form of the domain wall function $M(y)$ and strength $\kappa$) is given by: $D^{domain-wall}(V;x,y)\propto (\frac{L}{h v_{F}})^2 e^{-2 \kappa \int_{0}^{|y|}\,dy' M(y')}$.
This show the  similarity between  the result  obtain from the $domain-wall$ model and  the large numbers of  of dislocations given in equation $(54)$.

\vspace{0.1 in} 

\textbf{IVD-The tunneling  density of states $D^{(n=\pm1)}(V;\theta,u)$  for the $n=\pm1 $ contours.}

\vspace{0.1 in}

Following the  same procedure  as used for the $n=0$  and using  the eigenfunctions for $n=\pm1$ we  find :
\begin{equation}
D^{(n=\pm1)}(V;\theta,u)\equiv D^{(n=\pm1)}(\mu,V;\theta,u)_{even}+ D^{(n=\pm1)}(\mu,V;\theta,u,\mu)_{odd}
\label{contoursd}
\end{equation}
For the even $k$'s, we solve for the momentum $Q_{+}$ and  $Q_{-}$ and find:
\begin{eqnarray}
&&D^{(n=\pm1)}(\mu,V;\theta,u)_{even}=\frac{(B^{(2)})^{2}}{2\pi R_{g}(1) D(1) \sqrt{G(\theta,u)}}\sum_{Q_{r}=Q_{+},Q_{-}}\sum_{l=0}^{\infty}\delta[eV+\mu-\hbar v_{F}\sqrt{(\frac{l+\frac{1}{2}}{ R_{g}(1)})^2+Q_{r}^2}]\nonumber\\&&
[(e^{-2\zeta(\theta,u)\sin[2\theta]}+e^{2\zeta(\theta,u)\sin[2\theta]})((\sin[Q_{r}u-\frac{1}{2}\kappa(l,Q_{r})])^2+(\sin[Q_{r}u+\frac{1}{2}\kappa(l,Q_{r})])^2)+\nonumber\\&& 2(-1)^{l}\sin[Q_{r}u+\frac{1}{2}\kappa(l,Q_{r})]\sin[Q_{r}u-\frac{1}{2}\kappa(l,Q_{r})]\cdot\nonumber\\&&(
\cos[l(\theta+\frac{3\pi}{2}) -\zeta(\theta,u)(-\eta(u)+\cos[2\theta])]-\cos[(l+1)(\theta+\frac{3\pi}{2}) +\zeta(\theta,u)(-\eta(u)+\cos[2\theta])]];\nonumber\\&&
\end{eqnarray}
Similarly for the odd $k$'s we find:
\begin{eqnarray}
&&D^{(n=\pm1)}(\mu,V;\theta,u)_{odd}=\frac{(B^{(2)})^{2}}{2\pi R_{g}(1) D(1) \sqrt{G(\theta,u)}} \sum_{Q_{r}=Q_{+},Q_{-}}\sum_{l=0}^{\infty}\delta[eV+\mu-\hbar v_{F}\sqrt{(\frac{l+\frac{1}{2}}{ R_{g}(1)})^2+Q_{r}^2}]\nonumber\\&&
[(e^{-2\zeta(\theta,u)\sin[2\theta]}+e^{2\zeta(\theta,u)\sin[2\theta]})((\cos[Q_{r}u-\frac{1}{2}\kappa(l,Q_{r})])^2+(\cos[Q_{r}u+\frac{1}{2}\kappa(l,Q_{r})])^2)+\nonumber\\&& 2(-1)^{l}\cos[Q_{r}u+\frac{1}{2}\kappa(l,Q_{r})]\cos[Q_{r}u-\frac{1}{2}\kappa(l,Q_{r})]\cdot\nonumber\\&&(
\cos[l(\theta+\frac{3\pi}{2}) -\zeta(\theta,u)(-\eta(u)+\cos[2\theta])]-\cos[(l+1)(\theta+\frac{3\pi}{2}) +\zeta(\theta,u)(-\eta(u)+\cos[2\theta])]]\nonumber\\&&
\end{eqnarray}
For the present case the energy scale of the excitations is governed by the radius $R_{g}(1)$ and width $D$. The spectrum is discrete and we can't replace it by a continuum density of states as we did for the case $n=0$.

In figure $5$ we show the tunneling density of states at a fixed polar angle $\theta=\frac{\pi}{2}$  as a function of the voltage $V$. We observe that the density of states is dominated by high energy eigenvalues. This solutions are localized in energy. The range of the spectrum is above  $\mu+eV>200mV$ which  is well separated from  the low energy spectrum controlled by the $n=0$ contour (which ranges from $-120 mV $ to $70 mV$).

 Figure $6$   shows the  tunneling density of states as a function of the polar angle $\theta$ for a fixed energy . The periodicity in $\theta $  is controlled  by the discrete energy   eigenvalues. 

In figure  $7$  we show  the tunneling density of states  at a fixed voltage $V$  as a function of the polar angle $0<\theta<\pi$ and width $|u|<0.1 $.

\vspace{0.2 in}

\textbf{V-The  charge current-the in plane spin on the  surface of the $h^{T.I.}$ Hamiltonian}

\vspace{0.1 in}

\textbf{A-The   current in the absence of the edge dislocation for the  $h^{T.I.}$}

\vspace{0.1 in}

From the Hamiltonian given in equation $1$ we  compute the equation of motion for the velocity operator:
 $\frac{dx}{dt}=\frac{1}{i\hbar}[x,h]=v_{F}\sigma^{y}$ , $\frac{dy}{dt}=\frac{1}{i\hbar}[y,h]=-v_{F}\sigma^{x}$.   We  multiply the velocity operator  by the charge $(-e)$ and  identify the charge current operators :
$\hat{J}_{x}=(-e)v_{F}\sigma^{2}$,  $\hat{J}_{y}=(-e)(-v_{F})\sigma^{1}$.
This  also represent the "`real"' spin on the surface. Therefore, the charge current is a measure of the in-plane spin on the  surface.

Integrating over the $y$ coordinate we obtain the current $I_{x}^{T.I.}$ in the $x$ direction. Using the  eigenstates   $\Gamma_{p,q}(x,y)$  and $\Gamma_{-p,q}(x,y)$ of the $h^{T.I.}$ Hamiltonian 
\begin{equation*}
\Gamma_{p,q}(x,y)=e^{i p x}e^{iq y} \left(\begin{array}{cc} 1\\ i e^{i\chi(p,q)}\end{array}
\right)
\end{equation*}
\begin{equation*}
\Gamma_{-p,q}(x,y)=e^{-i p x}e^{iq y} \left(\begin{array}{cc} 1\\ -i e^{-i\chi(p,q)}\end{array}
\right)
\end{equation*}
we find $(\Gamma_{p,q}(x,y))(\sigma^{2})(\Gamma_{p,q}(x,y))=- (\Gamma_{-p,q}(x,y))(\sigma^{2})(\Gamma_{-p,q}(x,y))$ therefore, we conclude that the current $I_{x}^{T.I.}=0$ is zero.

\vspace{0.1 in}

\textbf{VB-The  current  in the presence of the edge dislocation}

\vspace{0.1 in}

We  will compute the current in the presence  of the edge dislocation.
The current operator $\hat{J}^{edge}_{x}(x,y)$  will be given in terms of the transformed currents. We find that the  current  density operator $J^{edge}_{x}(x,y)$ is given by:
\begin{equation}
\hat{J}^{edge}_{x}(x,y)=(-e)v_{F}[\sigma^{2}e^{x}_{1}-\sigma^{1}e^{x}_{2}]= (-e)v_{F}\sigma^{2}-(-e)v_{F}\frac{B^{(2)}}{2\pi}(\frac{ y\sigma^{1}+ x\sigma^{2}}{x^2+y^2})\approx (-e)v_{F}\sigma^{2}
\label{edge}
\end{equation}
We use the zero order  current operator  $\hat{J}^{edge}_{x}(x,y)\approx (-e)v_{F}\sigma^{2}$ to construct  the second quantization form for  the current  density. The operator  is defined with respect the to  shifted ground state $|\mu>\equiv|\tilde{0}>$  with the energy $E=\epsilon-\mu$ measured with respect the chemical potential and spinor  field $\Psi_{n=0}(x,y)$.
\begin{equation}
J^{edge}_{x}(x,y)= <\mu|\Psi_{n=0}^{\dagger}(x,y)\hat{J}^{edge}_{x}(x,y)\Psi_{n=0}(x,y)|\mu>
\label{spinor}
\end{equation} 
 Using the spinor eigenfunction given in equation $(35)$ and  the second quantized form with the electron like operators  $\alpha_{E,R}$,$\alpha_{E,L}$ and hole like $\beta_{E,R}$,$\beta_{E,L}$ we find :
\begin{equation*}
\Psi_{n=0}(x,y;t)\approx\sum_{E>0} [\alpha_{E,R}\left(\begin{array}{cc}U^{(n=0,R)}_{\uparrow}(x,y)\\U^{(n=0,R)}_{\downarrow}(x,y)\end{array}
\right)_{E+\mu}{e^{-i\frac{E}{\hbar}t}}+\beta^{\dagger}_{E,R} \left(\begin{array}{cc}U^{(n=0,R)}_{\uparrow}(x,y)\\U^{(n=0,R)}_{\downarrow}(x,y)\end{array}
\right)_{-E+\mu}e^{i\frac{E}{\hbar}t}
\end{equation*}
\begin{equation} +\alpha_{E,L}\left(\begin{array}{cc}U^{(n=0,L)}_{\uparrow}(x,y)\\U^{(n=0,L)}_{\downarrow}(x,y)\end{array}
\right)_{E+\mu}e^{-i\frac{E}{\hbar}t}+\beta^{\dagger}_{E,L} \left(\begin{array}{cc}U^{(n=0,L)}_{\uparrow}(x,y)\\U^{(n=0,L)}_{\downarrow}(x,y)\end{array}
\right)_{-E+\mu}e^{i\frac{E}{\hbar}t}]
\label{eqsp}
\end{equation} 
The current is a sum  of two terms computed with the eigen spinor obtained in equation $(35)$: $[U^{(n=0,R)}_{\uparrow}(x,y),U^{(n=0,R)}_{\downarrow}(x,y)]^{T}  \sigma^{2}[U^{(n=0,R)}_{\uparrow}(x,y),U^{(n=0,R)}_{\downarrow}(x,y)]$ 
 and  
 $[U^{(n=0,L)}_{\uparrow}(x,y),U^{(n=0,L)}_{\downarrow}(x,y)]^{T}  \sigma^{2}[U^{(n=0,L)}_{\uparrow}(x,y),U^{(n=0,L)}_{\downarrow}(x,y)]$ 
which have  opposite signs.  Due to the parity violation caused by the dislocation,   the density of states is asymmetric    $1\pm\frac{1}{\pi}\frac{hv_{F}}{L (\mu+V)}\sqrt{1-(\frac{\epsilon_{\bot} }{\mu+V})^2}\hspace{0.1 in})$ resulting in a finite current.  We 
integrate over the transversal direction  $y$ and obtain the  edge current  $I_{x}^{n=0,edge}$.
\begin{eqnarray}
&&I_{x}^{n=0,edge}= (-e)v_{F}\int_{-\frac{L}{2}}^ {\frac{L}{2}}\frac{d x}{L}\int_{-\frac{L}{2}}^{\frac{L}{2}}\,dy<\mu|J^{edge}_{x}(x,y)|\mu>=\nonumber\\&&
\frac{(-e)v_{F}}{4\pi}(\frac{L}{h v_{F}})^2(\frac{1}{L})\int_{-\frac{L}{2}}^ {\frac{L}{2}}\frac{d x}{L}\int_{-\frac{L}{2}}^{\frac{L}{2}}\frac{d y}{L}\frac{ e^{\frac{-B^{(2)}}{\pi}(\frac{x}{x^2+y^2+a^2})}}{\sqrt{G(x,y)}}
\int\,d\epsilon_{||}\int d\epsilon_{\bot}H[\mu-\sqrt{(\epsilon_{||})^2+(\epsilon_{\bot})^2}\hspace{0.1 in}]\frac{(hv_{F}/L)\cdot \epsilon_{||}}{(\epsilon_{||})^2+(\epsilon_{\bot})^2}\nonumber\\&&
=\frac{1}{4\pi}(\frac{-e v_{F}}{L})(\frac{\mu}{hv_{F}/L})f[\frac{B^{(2)}}{L}]\cdot(H[\mu+eV-\frac{hv_{F}}{L}]-H[\mu+eV-E_{max.}]);\hspace{0.2 in}
f[\frac{B^{(2)}}{L}]\approx 6.22\nonumber\\&&
\end{eqnarray}
$H[\mu-\sqrt{(\epsilon_{||})^2+(\epsilon_{\bot})^2}\hspace{0.1 in}]$ is the step function which is one for $\sqrt{(\epsilon_{||})^2+(\epsilon_{\bot})^2}\leq \mu$. The single particle energies are $\epsilon_{\bot}=\hbar v_{F}q_{\pm}$  and $\epsilon_{||}=\hbar v_{F}p$.
For $L\approx 10^{-6}m $, chemical potential  $\mu=120mV$   and $\frac{L}{B^{(2)}}\approx 100$ we find  that the current  $I_{x}^{n=0,edge}$ is in the range of $mA$.

To conclude, we have shown that the presence of an edge dislocation gives rise to a non-zero current which is a manifestation of the in-plane component of the spin on the two dimensional  surface . Therefore a  nonzero value  $I_{x}^{n=0,edge}\neq 0$ will be an indication of the presence of the edge dislocation. This  effect might be measured using  a coated tip with magnetic material used by the technique of Magnetic Force Microscopy.

\vspace{0.2 in}

\textbf{VI-Conclusions}

\vspace{0.2 in}

We have used the  coordinate transformation  method  to investigate $TI$ in the presence of deformations.  We have computed the spin connection and the metric tensor for the three dimensional $TI$.  This theory  has been applied to the surface  of a $TI$   with an edge dislocation. 
We have shown that the tunneling density of states is confined to  two dimensional region    $n=0$ and to high energy  circular contours  with $n=\pm1$.
 The edge dislocations violate the parity symmetry. As a result a current which is a manifestation of  in plane spin orientation is generated.
The in plane spin orientation is a manifestation of the  parity violation  induced by the  edge dislocation.
We propose that scanning tunneling methods   might be  able to   verify our prediction.

\pagebreak
  
\vspace{0.2 in}

\textbf{ Appendix -A}

\vspace{0.2 in}

We consider that    a two dimensional   manifold with  a   mapping  from the curved space   $X^{a}$, $ a=1,2 $,   to the $local $ $flat$ space   $x^{\mu}$,     $\mu=x,y$ exists.
We introduce the tangent vector \cite{Green}
 $e^{a}_{\mu}(\vec{x})=\frac{\partial X^{a}(\vec{x})}{\partial x^{\mu}}$,  $\mu=x,y$
 which satisfies the orthonormality relation  $e^{a}_{\mu}(\vec{x})e^{b}_{\mu}(\vec{x})=\delta_{a,b}$  (here we use the convention that we sum over indices which appear twice).  The metric tensor  for the curved space is given in terms of the flat metric $\delta_{a,b}$ and the scalar product of the tangent vectors: $e^{a}_{\mu}(\vec{x})e^{a}_{\nu}(\vec{x})=g_{\mu,\nu}(\vec{x})$.
The linear connection is determined by the Christoffel   tensor  $ \Gamma^{\lambda}_{\mu,\nu} $ :
  
\begin{equation}
 \nabla_{\partial_{\mu}}\partial_{\nu}= -\Gamma^{\lambda}_{\mu,\nu} \partial_{\lambda}
 \label{christ}
 \end{equation}
 
The Christoffel tensor is constructed from the metric tensor $ g_{\mu,\nu}(\vec{x})$.
\begin{equation}
\Gamma^{\lambda}_{\mu,\nu}=-\frac{1}{2}\sum_{\tau=x,y}g^{\lambda,\tau}(\vec{x})[\partial_{\nu}g_{\nu,\tau}(\vec{x})+\partial_{\mu}g_{\nu,\tau}(\vec{x})-\partial_{\tau}g_{\mu,\nu}(\vec{x})]
\label{gama}
\end{equation}

Next, we introduce the vector  field $\vec{V}=V^{a}\partial_{a}=V^{\mu}\partial_{\mu}$  where $a=1,2$  are the components in the curved space and $\mu=x,y$ represents the coordinate in the fixed cartesian  frame. The covariant derivative of the vector field  $V^{a}$ is determined by the  spin connection $\omega_{q,b}^{\mu}$ which needs to be computed: 

\begin{equation}
D_{\mu}V^{a}(\vec{x})=\partial_{\mu}V^{a}(\vec{x})+\omega_{a,b}^{\mu} V^{b}
\label{vector}
\end{equation}

For a two component spinor, we can identify the spin connection in the following way:   The spinor in the  the curved space (generated by the dislocation) is represented by $\widetilde{\Psi}(\vec{X})$  and  in the Cartesian space it is given by  is given  by  $\Psi(\vec{x})$   \cite{Maggiore}.
 The two component spinor represents a chiral fermion  which transform under spatial rotation as  spin half fermion:  

\begin{eqnarray}
&&\widetilde{\Psi}(\vec{X})=e^{\frac{-i}{2}\omega_{1,2}\sigma_{3}}\Psi(\vec{x})\nonumber\\&& e^{\frac{-i}{2}\omega_{1,2}\sigma_{3}}\equiv e^{\frac{1}{2}\omega_{a,b}\Sigma^{a,b}}\equiv e^{\sum_{a=1,2}\sum_{b=1,2}\frac{1}{2}\omega_{a,b}\Sigma^{a,b}}\nonumber\\&&
\omega_{a,b}\equiv-\omega_{b,a}\nonumber\\&&
\Sigma^{a,b}\equiv \frac{1}{4}[\sigma^{a},\sigma^{b}]\nonumber\\&&
\end{eqnarray}
 We have used the anti symmetric property of the rotation matrix  $\omega_{a,b}\equiv-\omega_{b,a}$, and the representation of the  generator $\Sigma^{a,b}$ in terms of the Pauli matrices. 
 
Therefore for a two component spinor we obtain the connection:

\begin{equation}
D_{\mu}\Psi(\vec{x})=(\partial_{\mu}+\frac{1}{2}\omega^{a,b}_{\mu}\Sigma_{a,b})\Psi(\vec{x})\equiv (\partial_{\mu}+\frac{1}{8}\omega^{a,b}_{\mu}[\sigma_{a},\sigma_{b}])\Psi(\vec{x})
\label{tensor}
\end{equation}

Next we will compute the spin connection  $\omega^{a,b}_{\mu}$ using the \textbf{Christoffel tensor}.
 In the physical coordinate  basis $x^{\mu}$  the covariant derivative  $D_{\mu}V^{\nu}(\vec{x})$ is determined by the Christoffel tensor: 

\begin{equation}
D_{\mu}V^{\nu}(\vec{x})=\partial_{\mu}V^{\nu}(\vec{x})+\Gamma^{\lambda}_{\mu,\nu}  V^{\lambda}
\label{connection}
\end{equation}

The relation between the spin connection and the linear connection can be obtained from the fact that the two covariant derivative of the vector $\vec{V}$ are equivalent.

\begin{equation}
D_{\mu}V^{a}=e^{a}_{\nu}D_{\mu}V^{\nu}
\label{rel}
\end{equation}

Since we have the relation $V^{a}=e^{a}_{\nu}V^{\nu}$  it follows from the last equation 

\begin{equation}
D_{\mu}[e^{a}_{\nu}]=D_{\mu}\partial_{\nu}e^{a}=(D_{\mu}\partial_{\nu})e^{a}+ \partial_{\nu}(D_{\mu}e^{a})=0
\label{relt}
\end{equation}

Using the definition of the Christoffel index  and the differential geometry relation $\nabla_{\partial_{\mu}}\partial_{\nu}= -\Gamma^{\lambda}_{\mu,\nu} \partial_{\lambda}$
\cite{Green},  we obtain  the relation between the spin connection and the linear connection:

\begin{equation}
D_{\mu}[e^{a}_{\nu}]=\partial_{\mu}e^{a}_{\nu}(\vec{x})-\Gamma^{\lambda}_{\mu,\nu} e^{a}_{\lambda}(\vec{x})+ \omega^{a}_{\mu,b}e^{b}_{\nu}(\vec{x})\equiv0
\label{dd}
\end{equation}

Solving this  equation, we obtain  the spin connection  given in terms of the Burger vector. 
We multiply from left equation  $(70)$ by the tangent  vector $e^{a}_{\nu}$ and  replace  $\Gamma^{\lambda}_{\mu,\nu}$ with the representation given in equation $(63)$.  We use the  metric tensor relations  $e^{a}_{\mu}(\vec{x})e^{b}_{\mu}(\vec{x})=\delta_{a,b}$,  $e^{a}_{\mu}(\vec{x})e^{a}_{\nu}(\vec{x})=g_{\mu,\nu}(\vec{x})$. 
and  find \cite{Green}:
\begin{eqnarray} &&\omega^{a,b}_{\mu}=\frac{1}{2}e^{\nu,a}(\partial_{\mu}e^{b}_{\nu}-\partial_{\nu}e^{b}_{\mu})-
\frac{1}{2}e^{\nu,b}(\partial_{\mu}e^{a}_{\nu}-\partial_{\nu}e^{a}_{\mu})\nonumber\\&&
-\frac{1}{2}e^{\rho,a}e^{\sigma,b}(\partial_{\rho}e_{\sigma, c}-\partial_{\sigma}e_{\rho, c})e^{c}_{\mu}
\end{eqnarray}
We notice the asymmetry   between $e^{\nu,a}$ and  $e_{a,\nu}$:
 
$e^{\nu,a}\equiv g^{\nu,\lambda}e^{a}_{\lambda}$ and  $e_{a,\nu}\equiv \delta_{a,b} e^{b}_{\nu}$

For our case we have  a two component the  spin connection $\omega_{x}^{1 2}$ and  $\omega_{y}^{1 2}$ 
\begin{eqnarray}
&&\omega_{x}^{1 2}=\frac{1}{2}e^{\nu,1}(\partial_{x}e^{2}_{\nu}-\partial_{\nu}e^{2}_{x})
-\frac{1}{2}e^{\nu,2}(\partial_{x}e^{1}_{\nu}-\partial_{\nu}e^{1}_{x})
-\frac{1}{2}e^{\rho,a}e^{\sigma,b}(\partial_{\rho}e_{\sigma,c}-\partial_{\sigma}e_{\rho,c})e^{c}_{x};\nonumber\\&&
\omega_{y}^{1 2}=\frac{1}{2}e^{\nu,1}(\partial_{y}e^{2}_{\nu}-\partial_{\nu}e^{2}_{y})
-\frac{1}{2}e^{\nu,2}(\partial_{y}e^{1}_{\nu}-\partial_{\nu}e^{1}_{y})
-\frac{1}{2}e^{\rho,a}e^{\sigma,b}(\partial_{\rho}e_{\sigma,c}-\partial_{\sigma}e_{\rho,c})e^{c}_{y}\nonumber\\&&
\end{eqnarray}
These equations are  further simplified with the help of  equations $(11-17)$  with $e^{1}_{y}=0$ , $e^{1}_{x}=1$ and the Burger    tensor  $\partial_{x}e^{2}_{y}-\partial_{y}e^{2}_{x}=B^{(2)}\delta^{2}(\vec{r})$ . 
\begin{eqnarray}
&&\omega_{x}^{1 2}=\frac{1}{2}g^{\nu,\lambda}e^{1}_{\lambda}(\partial_{x}e^{2}_{\nu}-\partial_{\nu}e^{2}_{x})-\frac{1}{2}g^{\rho,r}e^{1}_{r}g^{\rho,s}e^{2}_{s}[\partial_{\rho}( \delta_{c,b}e^{b}_{\sigma})-\partial_{\sigma}( \delta_{c,d}e^{d}_{\rho})]e^{c}_{x}=\nonumber\\&&
\frac{1}{2}B^{(2)}\delta^{(2)}(\vec{r})[g^{y,x}e^{1}_{x}+g^{y,y}e^{1}_{y}-(g^{x,r}g^{y,s}-
g^{y,r}g^{x,s})(e^{1}_{r}e^{2}_{s}e^{2}_{x})]=\nonumber\\&&
\frac{1}{2}B^{(2)}\delta^{(2)}(\vec{r})[g^{y,x}e^{1}_{x}-(g^{x,x}g^{y,y}-
g^{y,x}g^{x,y})e^{1}_{x}e^{2}_{y}e^{2}_{x}]\approx\nonumber\\&&\frac{1}{2}B^{(2)}\delta^{(2)}(\vec{r})[- \frac{B^{(2)}}{2\pi}\frac{y}{y^2+x^2} -(1-(\frac{B^{(2)}}{2\pi}\frac{y}{y^2+x^2})^2)(\frac{B^{(2)}}{2\pi}\frac{y}{x^2+y^2})(1-\frac{B^{(2)}}{2\pi}\frac{x}{x^2+y^2})]\approx \nonumber\\&&
\frac{1}{2}B^{(2)}\delta^{(2)}(\vec{r})[-\frac{B^{(2)}}{2\pi}\frac{2y-x}{y^2+x^2}]\nonumber\\&&
\end{eqnarray}
and
\begin{eqnarray}
&&\omega_{y}^{1 2}=\frac{1}{2}e^{\nu,1}(\partial_{y}e^{2}_{\nu}-\partial_{\nu}e^{2}_{y})
-\frac{1}{2}e^{\nu,2}(\partial_{y}e^{1}_{\nu}-\partial_{\nu}e^{1}_{y})
-\frac{1}{2}e^{\rho,1}e^{\sigma,2}[\partial_{\rho}(\delta_{c,b}e_{\sigma}^{b})-\partial_{\sigma}(\delta_{c,d}e_{\rho}^{d})]e^{c}_{y}=\nonumber\\&&
\frac{1}{2}g^{\nu,\lambda}e^{1}_{\lambda}[\partial_{y}e^{2}_{\nu}-\partial_{\nu}e^{2}_{y}]
-\frac{1}{2}g^{\nu,r}e^{1}_{r}[\partial_{y}e^{1}_{\nu}-\partial_{\nu}e^{1}_{y}]-\frac{1}{2}g^{\rho,r}e^{1}_{r}g^{\sigma,s}e^{2}_{s}[\partial_{\rho}e^{c}_{\sigma}-\partial_{\sigma}e^{c}_{\rho}]e^{c}_{y}=\nonumber\\&&
-\frac{B^{(2)}}{2}\delta^{(2)}(\vec{r})g^{x,\lambda}e^{1}_{\lambda}-\frac{B^{(2)}}{2}\delta^{(2)}(\vec{r})[g^{x,r}g^{y,s}-g^{y,r}g^{x,s}]e^{1}_{r}e^{2}_{s}e^{2}_{y}\approx -\frac{B^{(2)}}{2}\delta^{(2)}(\vec{r}) \nonumber\\&&
\end{eqnarray}
To first order first  the Burger vector $B^{(2)}$ the spin connections are given by :
$\omega_{x}^{1 2}=-\omega_{x}^{21}\approx 0$   and  $\omega_{y}^{1 2}=- \omega_{y}^{2 1}\approx -\frac{1}{2}B^{(2)}\delta^2(\vec{r})$.










\pagebreak


\clearpage
\begin{figure}
\begin{center}
\includegraphics[width=7.0 in ]{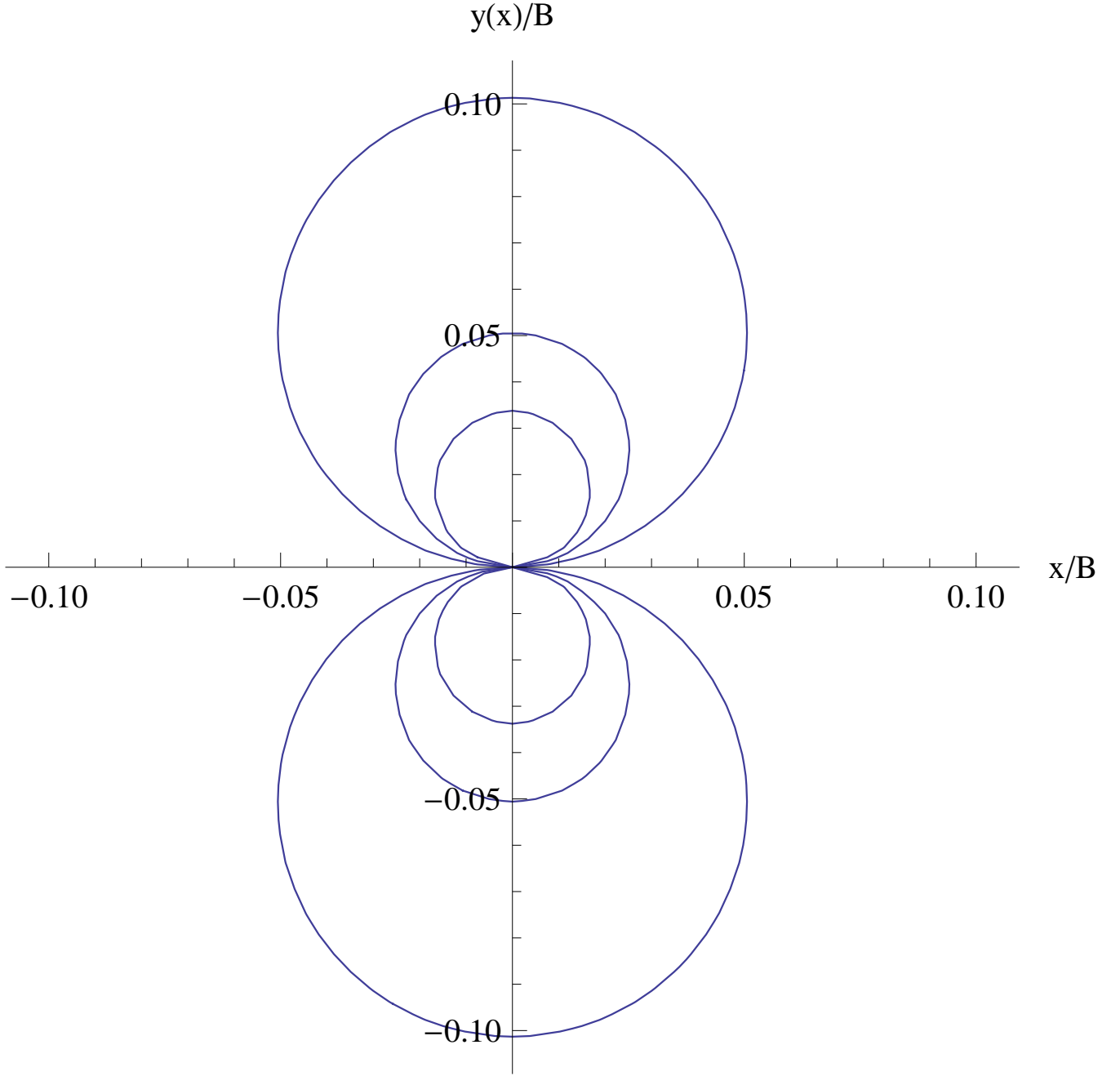}
\end{center}
\caption{The contours  $(x(s))^2+ (y(s)-\frac{R_{g}}{ n})^2 =(\frac{R_{g}}{n})^2$ for $n=\pm1,\pm2,\pm3$(in  decreasing  size ),$R_{g}(n)=\frac{R_{g}}{n}$. $n=0$ corresponds to the equation $y(s)=0$ and $|y|>d $ (see the text).  The  the distance is measured in  units  of the Burger vector $B^{(2)}$.}   
\end{figure}



\clearpage



\clearpage
\begin{figure}
\begin{center}
\includegraphics[width=7.0 in ]{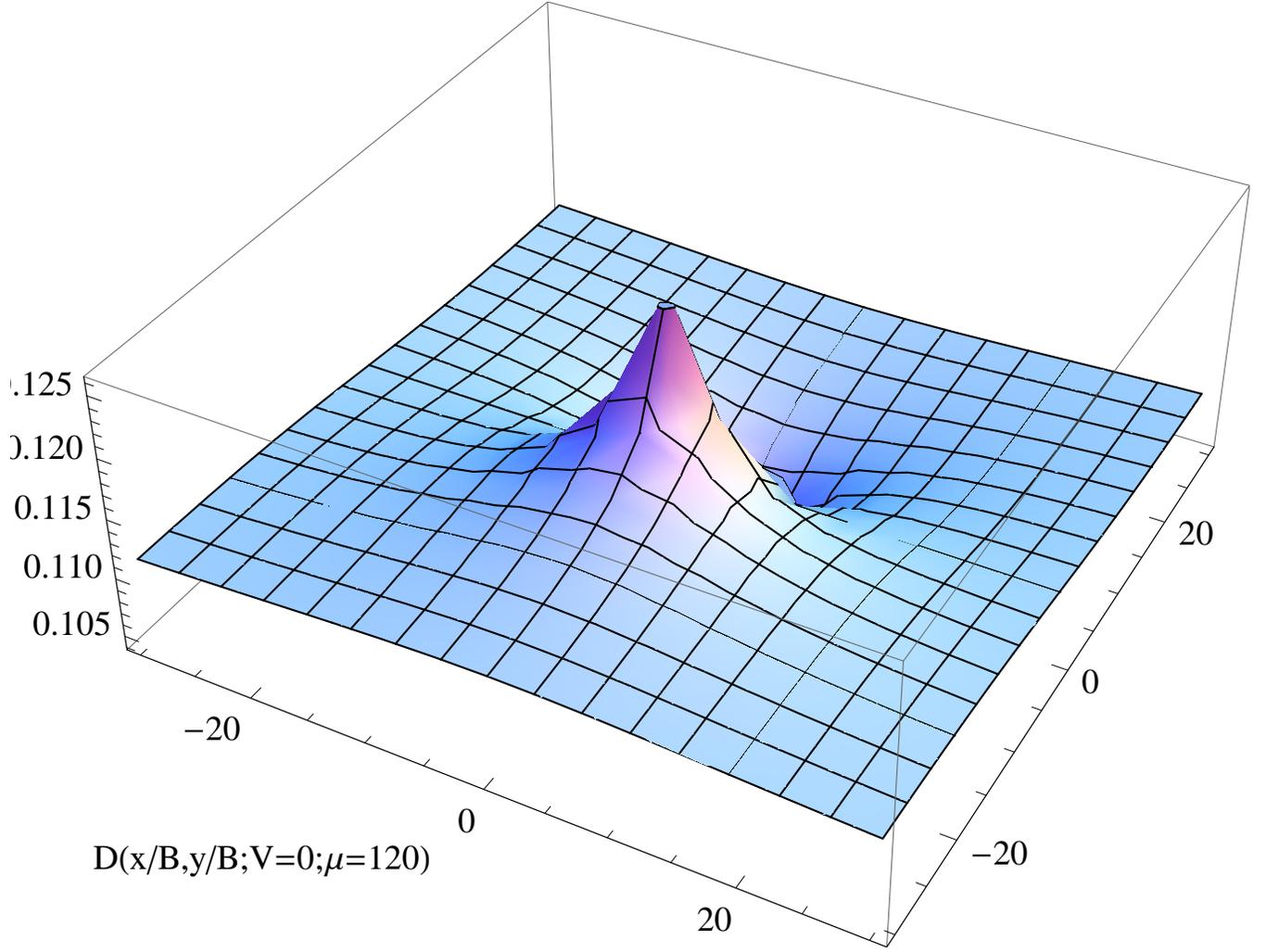}
\end{center}
\caption{The  tunneling density   of states  for  $n=0$  ,  $\frac{dI}{dV}\propto D^{(n=0)}(\frac{x}{B^{(2)}},\frac{y}{B^{(2)}};\mu=120mV) $.  The right corner  represents the intersection of the $x$ coordinate which runs from $30$ (right corner)  to $-30$  and the $y$ coordinate  which runs from $-30$ (right corner)  to $30$ in units of the Burger vector.}
\end{figure}

\clearpage

\begin{figure}
\begin{center}
\includegraphics[width=7.0 in ]{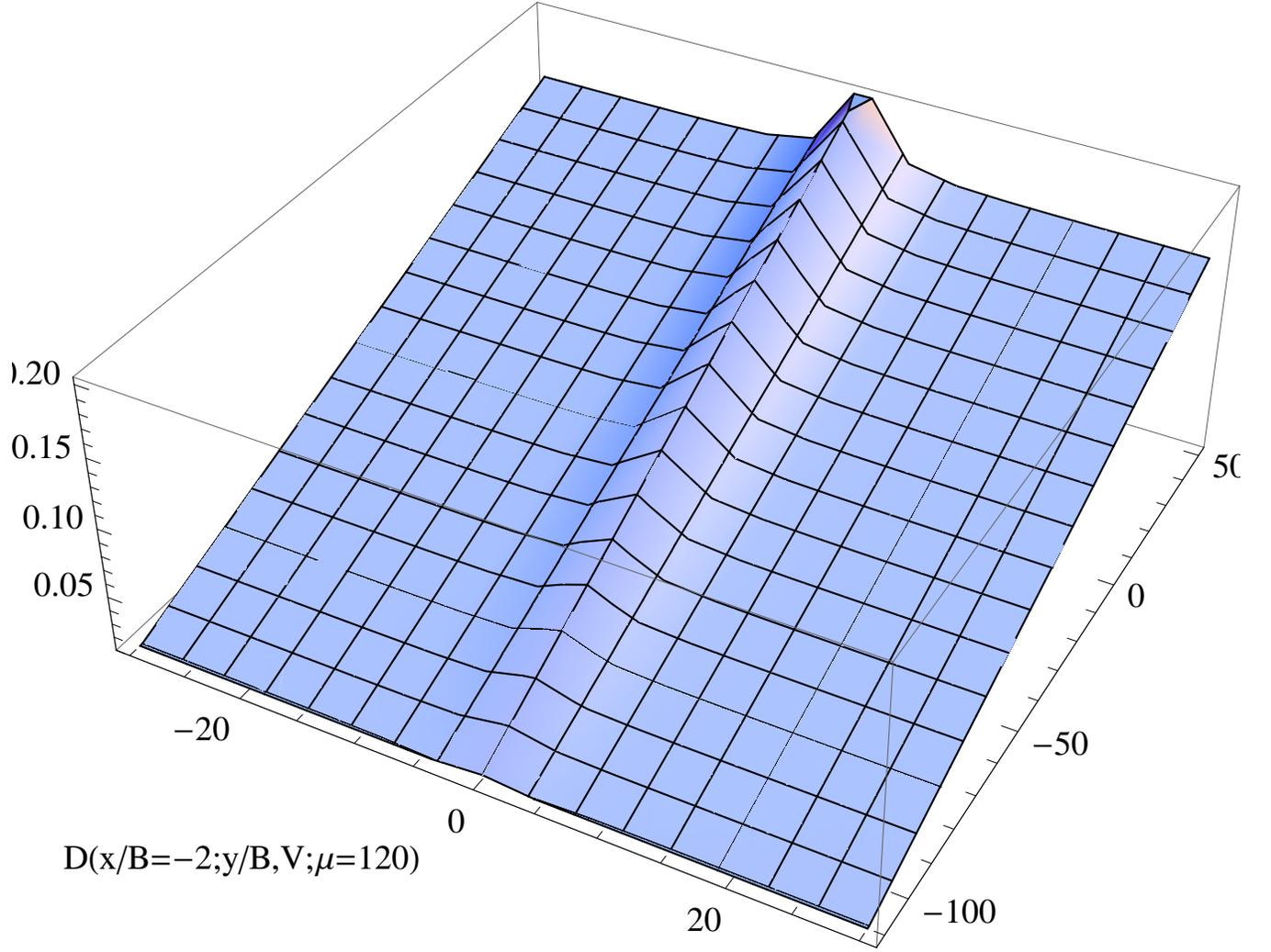}
\end{center}
\caption{The  tunneling density   of states  for  $n=0$  as a function  of $y$ and $V$ $\frac{dI}{dV}\propto D^{(n=0)}(\frac{x}{B^{(2)}}=-2,\frac{y}{B^{(2)}};\mu=120mV) $. The voltage range is $-120 \leq V  \leq  50$ and the $y$ coordinate is in the range   $-30 \leq \frac{y}{B^{(2)}}  \leq  30$.}
\end{figure}

\clearpage
\clearpage
\begin{figure}
\begin{center}
\includegraphics[width=7.0 in ]{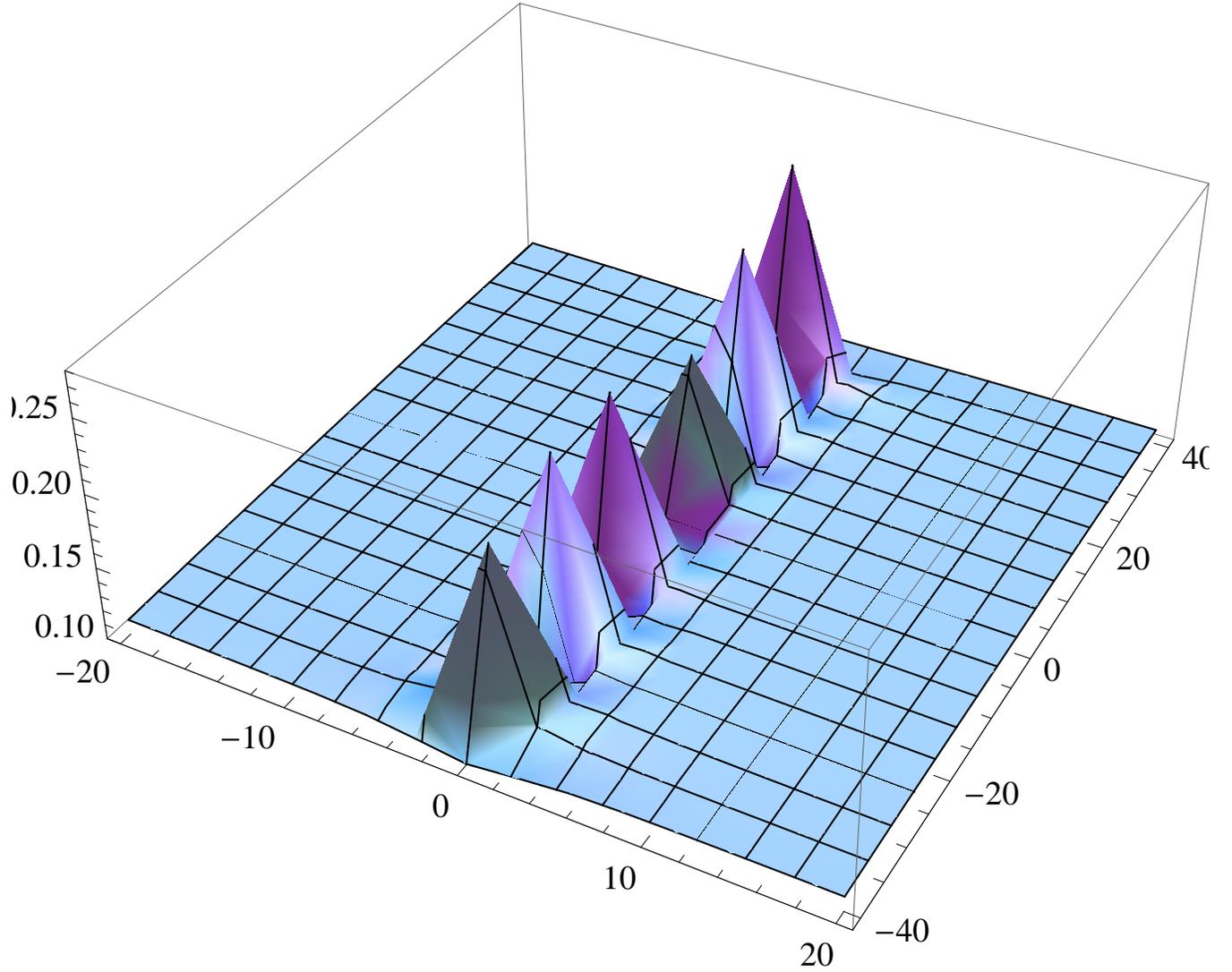}
\end{center}
\caption{Many Dislocations - with the core of the dislocations at   $[x_{w},y=0]$ , $w=1,2...2M$;   The maximum of the tunneling density of states  is confined along  $y=0$. The coordinates of the tunneling density of states are restricted to :  $-40 \leq \frac{x}{B^{(2)}} \leq  40$ and  $-20 \leq \frac{y}{B^{(2)}} \leq  20$. }  
\end{figure}
\clearpage

\clearpage
\begin{figure}
\begin{center}
\includegraphics[width=7.0 in ]{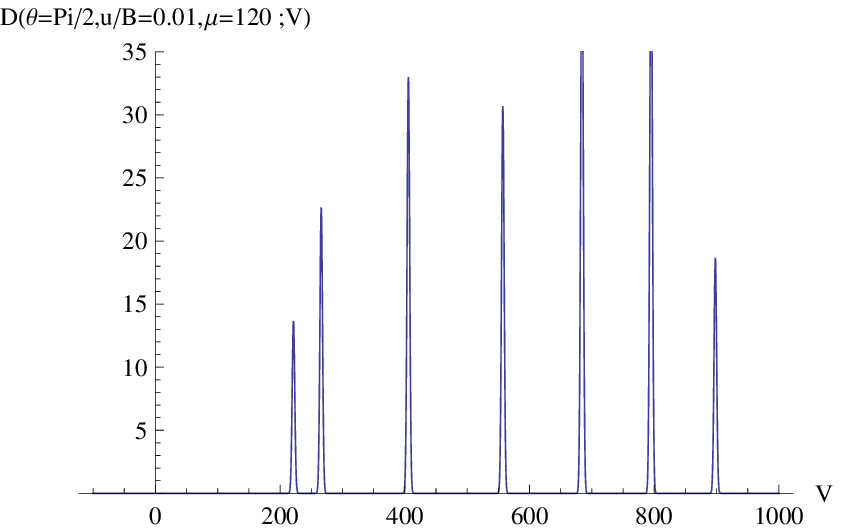}
\end{center}
\caption{The  discrete tunneling density  of states  for  $n=1$, as a function of the voltage $V$ $D^{(n=1)}(V;\theta =\frac{\pi}{2},\frac{u}{B^{(2)}},\mu=120mV)$}
\end{figure}
\clearpage

\begin{figure}
\begin{center}
\includegraphics[width=7.0 in ]{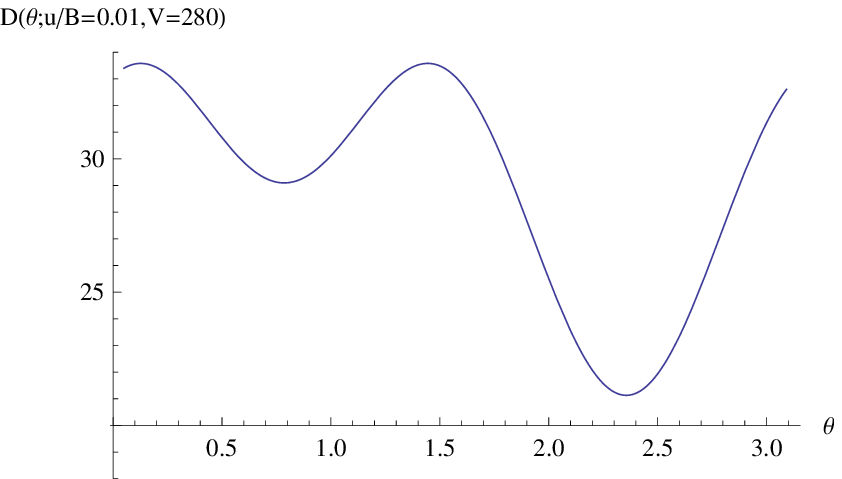}
\end{center}
\caption{The tunneling density of states as a function of $\theta$  $D^{(n=1)}(\theta; \frac{u}{B^{(2)}}=0.01, V=280 mV,\mu=120mV)$}  
\end{figure}

\clearpage

\clearpage
\begin{figure}
\begin{center}
\includegraphics[width=7.0 in ]{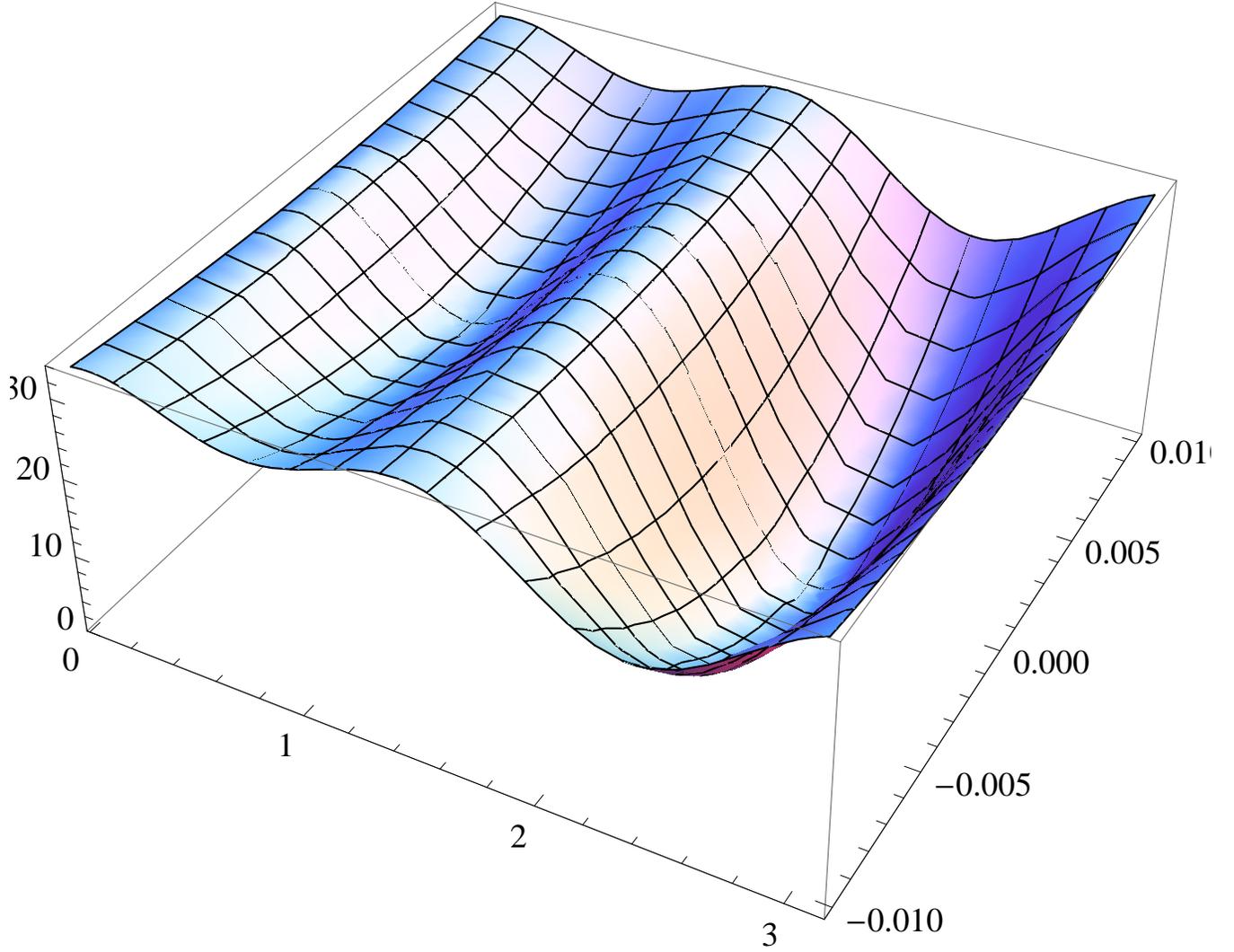}
\end{center}
\caption{The tunneling  density of states as a function of  $\theta$ and $u$ at a fixed voltage $V=280mV$   $D^{(n=1)}(\theta, \frac{u}{B^{(2)}}; V=280 mV,\mu=120mV)$}  
\end{figure}
\clearpage


\pagebreak


\end{document}